\documentclass[aps,pra,superscriptaddress,reprint, floatfix]{revtex4-1}
\usepackage{amstext,amssymb}
\usepackage{amsmath}
\usepackage{amsfonts}
\usepackage{xcolor}
\usepackage{graphicx}
\usepackage{bbold}
\usepackage[T1]{fontenc}
\usepackage{mathtools}
\DeclareMathOperator{\Tr}{Tr}
\newcommand{\bra}[1]{\left\langle #1 \right\vert}
\newcommand{\ket}[1]{\left\vert #1 \right\rangle}
\newcommand{\expect}[1]{\left\langle #1 \right\rangle}
\newcommand{\approxtext}[1]{\ensuremath{\stackrel{\text{#1}}{\approx}}}
\usepackage{tikz}
\usepackage{lipsum}

\begin{document}
\title{Quantum transient heat transport in the hyper-parametric oscillator}
\author{JungYun Han}
\affiliation{Center for Theoretical Physics of Complex Systems, Institute for Basic Science (IBS), Daejeon 34126, Republic of Korea.}
 \affiliation{Basic Science Program, University of Science and Technology (UST), Daejeon 34113, Republic of Korea.}
 \author{Daniel Leykam}
 \affiliation{Centre for Quantum Technologies, National University of Singapore, 3 Science Drive 2, Singapore 117543.}
\author{Dimitris G. Angelakis}
 \affiliation{Centre for Quantum Technologies, National University of Singapore, 3 Science Drive 2, Singapore 117543.}
 \affiliation{School of Electrical and Computer Engineering, Technical University of Crete, Chania, Crete 73100, Greece.}
 \author{Juzar Thingna}
 \email{jythingna@gmail.com}
\affiliation{Center for Theoretical Physics of Complex Systems, Institute for Basic Science (IBS), Daejeon 34126, Republic of Korea.}
  \affiliation{Basic Science Program, University of Science and Technology (UST), Daejeon 34113, Republic of Korea.}
\begin{abstract}
We explore nonequilibrium quantum heat transport in nonlinear bosonic systems in the presence of a non-Kerr-type interaction governed by hyper-parametric oscillation due to two-photon hopping between the two cavities. We estimate the thermodynamic response analytically by constructing the $su(2)$ algebra of the nonlinear Hamiltonian and predict that the system exhibits a negative excitation mode. Consequently, this specific form of interaction enables the cooling of the system by inducing a ground state transition when the number of particles increases, even though the interaction strength is small. We demonstrate a transition of the heat current numerically in the presence of symmetric coupling between the system and bath and show long relaxation times in the cooling phase. We compare with the Kerr-type Bose-Hubbard form of interaction induced via cross-phase modulation, which does not exhibit any such transition. We further compute the nonequilibrium heat current in the presence of two baths at different temperatures and observe that the cooling effect for the non-Kerr-type interaction persists. Our findings may help in the manipulation of quantum states using the system's interactions to induce cooling.
\end{abstract}
\maketitle

\section{Introduction}
Quantum photonics is the field of manipulating quantum states of light in integrated photonic circuits~\cite{Wang2019}, including the emulation of interesting electronic phenomena such as topological phases~\cite{Ozawa19}, exploration of non-Hermitian effects~\cite{Gong18}, and the design of novel integrated light sources~\cite{Harari18,Miguel18}. Furthermore, quantum photonic circuits form a promising platform for quantum computation and information processing, particularly for the storage, protection, and transmission of quantum information~\cite{Rechtsman16,Mittal16E,Tambasco18,Clemens19,Yuan19,Han20}.

Nonlinear quantum photonics pursues the emulation of novel phenomena in strongly correlated electronic systems and explores their peculiar behavior due to the bosonic statistics~\cite{Chang14,Noh_2016,Smirnova20}. The strong photon-photon interactions can be mediated via atoms in optical cavities (cavity-QED) and the microwave regime using superconducting circuits (circuit-QED)~\cite{Roushan1175,Blais20}. In the presence of inversion symmetry, the lowest order of nonlinearity is typically the third-order. For instance, Kerr-type interactions yield an intensity-dependent refractive index, enabling self-phase modulation (SPM) and cross-phase modulation (XPM). These interactions mimic the well-studied fermionic Hubbard interaction~\cite{altland10,XuPRB17} in atomic systems~\cite{Greiner02PT,Greiner02}. There are also non-Kerr-type interactions such as spontaneous four-wave mixing, in which two photons (provided by an intense pump beam) create two correlated photons in different modes (signal and idler).  Another non-Kerr-type interaction involves the pump-induced hopping of photon pairs between two cavities by creating the photons in the same mode, called two-photon hopping (TPH)~\cite{StepanenkoPRA20}. These effects form the basis for hyper-parametric oscillators~\cite{Razzari2010}, where different modes are coupled via third-order nonlinearity. These interactions exist only in bosonic systems due to the nature of the statistics and can be emulated using BECs~\cite{Krachmalnicoff10} or trapped ions~\cite{Ding2017,Ding2017b}.

Despite the utmost experimental protection, these systems always interact with an environment. Therefore, in this work, we study the heat exchange between the strongly nonlinear (interacting) quantum photonic system and its thermal environment. We show that Kerr-type interactions always cause the system to heat~\cite{Skelt19,Roberts20,Zhang_2013}, whereas non-Kerr-type interactions can lead to cooling. Specifically, in order to cool the system, a dual effect of the presence of a negative excitation mode (isolated system property) and an environment that can excite such a mode (system-bath property) should exist. The positive operator form of the Kerr-type interactions prohibits the existence of negative excitation mode and hence always causes the system to heat. We elucidate our main idea analytically using the \emph{deformed} $su(2)$ algebra approach, which is non-perturbative in the strength of nonlinearity and further corroborate the results using numerical simulations based on exact diagonalization of a truncated Hilbert space. In the case of two-photon hopping interactions, we compute the transient heat current in a coupled dimer and observe a net cooling effect when the negative excitation mode induces a new ground state. This allows the system to cool down, and we observe that this cooling is more efficient as we increase the interaction strength of TPH interaction and the number of photons in the system.

This paper's structure is as follows: We first introduce the quantum nonlinear Hamiltonian in the second quantized form and compare the different types of interaction terms by constructing the $su(2)$ algebra for the nonlinear Hamiltonians in Sec.~\ref{sec:quantum nonlinear}. In Sec.~\ref{sec: linear thermo}, we compute the transient heat current for the linearly coupled dimer analytically. The difference in the heat current between nonlinear interactions, TPH and XPM, is investigated in Sec.~\ref{sec: nonlinear thermo}. In both cases, we consider systems coupled to either one or two baths and study the transient and steady state heat currents, focusing on the different behaviors resulting from the linear and nonlinear interactions. We present our conclusions and future outlook in Sec.~\ref{sec:conclusion}.

\section{Quantum nonlinear photonic interactions}
\label{sec:quantum nonlinear}

We consider a model of optical nonlinearity wherein the system is a dimer of either cavities or ring resonators~\cite{Helt10,Sripakdee11,Vernon15,Kowalewska19} whose Hamiltonian is given by
\begin{equation}
\label{eq:Htot}
\begin{aligned}
    &H =  H_{\text{site}} + H_{\text{interaction}}, \quad 
    H_{\text{site}} = \omega (a^{\dagger}a + b^{\dagger}b), \\& H_{\text{interaction}} = \begin{cases} H_{\text{hop}} =  J\left(a^{\dagger}b + b^{\dagger}a\right),
   \\H_{\text{TPH}} = Y\left(a^{\dagger}a^{\dagger}bb + b^{\dagger}b^{\dagger}aa\right), \\
   H_{\text{XPM}} =  Z\left(a^{\dagger}ab^{\dagger}b\right).
   \end{cases}
\end{aligned}
\end{equation}
Above $a$ and $b$ are the bosonic field operators of the two cavities, each having the same frequency $\omega$. The mediating link between them could either be linear $H_{\text{hop}}$ with strength $J$ or nonlinear $H_{\text{TPH}}$ ($H_{\text{XPM}}$) with strength $Y$ ($Z$). 

The nonlinear interaction strengths $Y,Z \propto \omega^{2}$, where $\omega$ is the real photon frequency~\cite{Vernon15}. The two-photon hopping (TPH) is a non-Kerr-type interaction that only bosons can exhibit, and it induces two-photon oscillations~\cite{Helt10, Sripakdee11, Vernon15, Kowalewska19, Menotti19}. The second nonlinear interaction is the Kerr-type interaction called cross-phase modulation (XPM), in which the refractive index depends on the intensity of the photon in the other cavity~\cite{boyd03,Vernon15}. Cross-phase modulation is observed in both bosonic and fermionic systems, and it is often referred to as a Hubbard-type interaction~\cite{altland10}. In other words, TPH annihilates two photons at the same cavity simultaneously and creates two photons at a different cavity. In comparison, XPM destroys two photons from different cavities and creates `new' photons back at the same cavities.

Spectrum-wise, as $H_{\text{site}}$ and $H_{\text{XPM}}$ are positive operators, adding more photons always increases the system energy. On the other hand, hopping interactions $H_{\text{hop}}$ and $H_{\text{TPH}}$ are not positive operators, which could reduce the system energy by adding photons. To explicitly show this, we construct the $su(2)$ algebra of both the linear and nonlinear Hamiltonians. To be self-contained, we first define basis operators of the standard $su(2)$ algebra as,
\begin{equation}
\label{eq:stdsu2ops}
\begin{aligned}
    &X_{0} \coloneqq \frac{1}{2}\left(a^{\dagger}a + b^{\dagger}b\right), \quad X_{z} \coloneqq  \frac{1}{2}\left(a^{\dagger}a - b^{\dagger}b\right),   \\&X_{+} \coloneqq  a^{\dagger}b, \quad    X_{-} \coloneqq X_{+}^{\dagger},\\
    &X^{2} \coloneqq  \frac{X_{+}X_{-}+X_{-}X_{+}}{2} +X_{z}^{2},
\end{aligned}
\end{equation}
which satisfy
\begin{equation}
\begin{aligned}
    &[X_{z},X_{\pm}] = \pm X_{\pm}, \quad [X_{0},X_{\pm,z}] = 0, \\
    &[X_{+},X_{-}] = 2X_{z}, \quad X^{2} = X^{2}_{0} + X_{0},
\label{eq: algebra construction}
\end{aligned}
\end{equation}
where $X_{0}$ behaves like the identity operator and $X_{\pm}$ are the raising/lowering operators. $X_{z}$ represents the imbalance between sites and provides us with a reference axis. The operator $X^{2}$ is the Casimir operator with eigenvalue $(n/2)^{2} + (n/2)$, where $n$ is the total number of photons in the system, which is conserved. We then can construct a basis~\cite{Gottfried03} such that, 
\begin{eqnarray}
    X_{z}\ket{\frac{n}{2},m} &=& m\ket{\frac{n}{2},m}, \nonumber \\
    \label{eq:stdsu2ops2}
    X^{2}\ket{\frac{n}{2},m} &=& f_+\left(\frac{n}{2}\right)\ket{\frac{n}{2},m}, \\
    X_{\pm}\ket{\frac{n}{2},m} &=& \sqrt{f_+\left(\frac{n}{2}\right)- f_\pm\left(m\right)}\ket{\frac{n}{2},m \pm 1},\nonumber
\end{eqnarray}
where $f_\pm(x) = x(x\pm1)$, the first index $n/2$ determines the eigenvalue of $X_{0}$, analogous to the angular quantum number, and the eigenvalue of $X_{z}$ $m \in \lbrace -\frac{n}{2}, -\frac{n}{2}+1, ... ,\frac{n}{2}-1,\frac{n}{2} \rbrace$ is analogous to the magnetic quantum number. The corresponding vector space is a subspace of Fock space $\ket{n_{a},n_{b}}_{a,b}$. For example, for $n = 2$ the basis $\ket{1,m}$ can be mapped to the Fock state for cavity ($a$,$b$), i.e., $\ket{n_{a},n_{b}}_{a,b}$ with $n_{\alpha}$ being the number of photons at site $\alpha$ as,
\begin{equation}
    \ket{1,1} = \ket{2,0}_{a,b}, \ \ket{1,0} = \ket{1,1}_{a,b},\  \ket{1,-1} = \ket{0,2}_{a,b}. 
\end{equation}
Thus, the linear Hamiltonian ($H_{\rm site} + H_{\rm hop}$) in terms of the new operators defined above takes the form
\begin{equation}
    H = 2\omega X_{0} + 2J X_{x},
\end{equation}
where $2X_{x} \coloneqq X_{+} + X_{-}$ and $2X_{y} \coloneqq -i(X_{+} - X_{-})$. The operators defined in Eq.~\eqref{eq:stdsu2ops2} can also be \emph{uniquely} represented using the vector of operators $A^{\dagger}=\left( \begin{array}{cc}
		 a^{\dagger} &  b^{\dagger}
		\end{array} \right)$, i.e., $2X_{\alpha} = A^{\dagger}\, \sigma_\alpha \, A$ with $\alpha =\{0,x,y,z\}$, $\sigma_0$ being the $2\times 2$ identity matrix, and the other $\sigma$' representing the Pauli matrices. Thus, the linear Hamiltonian in terms of the vector of operators $A$ takes the form,
\begin{equation}
   H  = A^{\dagger} \left\{\omega \sigma_0 + J \sigma_{x} \right\} A.
\end{equation}
The above form could have been trivially expected since our Hamiltonian is quadratic in this case. The above $2\times2$ matrix (within curly brackets) can be diagonalized using the normal unitary transformation to define the normal modes,
\begin{equation}
    \left( \begin{array}{c}
		 c \\  d
		\end{array} \right) =\frac{1}{\sqrt{2}}\left( \begin{array}{cc}
		1 & 1 \\ 1 & -1
		\end{array} \right)\left( \begin{array}{c}
		a \\ b
		\end{array} \right).
\label{eq: Bogoliubov single}
\end{equation}
Here we interpret eigenvalues of the above matrix $\nu$ as the energy added by adding a single excitation to one of the normal modes, $E^{n}_{c/d} - E^{n-1}_{c/d} \coloneqq \nu$, where $E^{n}_{c/d}$ refers to the energy eigenvalue of $n$ photons in the mode $c/d$. The two basis operators are orthogonal and commute, i.e., $\Tr[c^{\dagger}d] = 0$ and $[c,d] = 0$. Thus, we obtain a completely decoupled spectrum with two normal modes, symmetric $(c)$ and anti-symmetric $(d)$ modes, with $\nu = \omega \pm J$. We note that for the case of coupled optical resonators, the linear (non-interacting) Hamiltonian employed is valid for weakly coupled sites, i.e., $\omega \gg J$, such that both normal modes always have positive energy. However, using state-of-the-art circuit QED or ion trap setups can to achieve the ultra-strong coupling regime with $J \sim \omega$~\cite{Bosman17, FriskKockum19}.

In the presence of the nonlinear TPH interaction, a similar construction is available, using the well-known Higgs method~\cite{Higgs79,Sklyanin82,Karassiov94,Klimov02}, by introducing the operators,
\begin{equation}
\begin{aligned}
    &\mathcal{Y}_{0} \coloneqq \frac{1}{2}X_{0}, \quad \mathcal{Y}_{z} \coloneqq \frac{1}{2}X_{z}, \quad  \mathcal{Y}_{+} \coloneqq  X^{2}_{+}, \quad    \mathcal{Y}_{-} \coloneqq \mathcal{Y}_{+}^{\dagger}.
\end{aligned}
\end{equation}
Above operators $\mathcal{Y}$ satisfy the commutation relations given in Eq.~\eqref{eq: algebra construction} with $X$ replaced by $\mathcal{Y}$ except for $[\mathcal{Y}_{+},\mathcal{Y}_{-}]$. In order to obtain the $\mathcal{Y}_{\pm}$ commutation relation we introduce the structure functional $\Phi(f_{\pm}(x))$ that represents the characteristic of the algebra~\cite{BONATSOS99},
\begin{equation}
\begin{aligned}
    &\mathcal{Y}^{2} = \Phi\left(f_+(\mathcal{Y}_{0})\right),
    \\& \mathcal{Y}_{+}\mathcal{Y}_{-} = \mathcal{Y}^{2}- \Phi(f_-(\mathcal{Y}_{z})), 
    \\& \mathcal{Y}_{-}\mathcal{Y}_{+} = \mathcal{Y}^{2} - \Phi(f_+(\mathcal{Y}_{z})), 
\end{aligned}
\end{equation}
where Casimir operator $\mathcal{Y}^2 = \mathcal{Y}_{+}\mathcal{Y}_{-} + \Phi(f_-(\mathcal{Y}_{z})) = \mathcal{Y}_{-}\mathcal{Y}_{+} + \Phi(f_+(\mathcal{Y}_{z}))$ commutes with all the basis operators $\mathcal{Y}_{x,y,z}$. The structure functional $\Phi(f_{\pm}(x))$ is a polynomial functional of $f_{\pm}(x)$, defined below Eq.~\eqref{eq:stdsu2ops2}, which ensures: i) the eigenvalue of Casimir operator $\mathcal{Y}^{2}$ is semi-positive, ii) the maximum eigenvalue of $\mathcal{Y}_{z}$ is $n/4$ (eigenvalue of $\mathcal{Y}_{0}$), iii) and the minimum is $-n/4$. Standard $su(2)$ algebra yields the structure functional $\Phi(f_{\pm}(x)) = f_{\pm}(x)$, but in our case we obtain the functional form of $\Phi(f_{\pm}(x))$ from the commutation relation,
\begin{eqnarray}
    [\mathcal{Y}_{+},\mathcal{Y}_{-}] &\coloneqq& \Phi(f_+(\mathcal{Y}_{z}))-\Phi(f_-(\mathcal{Y}_{z})) \nonumber \\
    &=& -64\mathcal{Y}^{3}_{z}+8(8\mathcal{Y}^{2}_{0}+4\mathcal{Y}_{0}-1)\mathcal{Y}_{z}.
\label{Eq: deformed commutator relation}
\end{eqnarray}
Thus, the structure functional of this \emph{deformed} $su(2)$ algebra takes the form
\begin{equation}
    \Phi(f_{\pm}(x)) = 4(8\mathcal{Y}^{2}_{0}+4\mathcal{Y}_{0}-1)f_{\pm}(x) - 16(f_{\pm}(x))^2.
\end{equation}
The action of the raising, lowering, and Casimir operators on the basis of the linear system $X_z$ yields polynomial coefficients in $n,m$~\cite{Bonatsos95,BONATSOS99}, i.e.,
\begin{eqnarray}
    \mathcal{Y}_{z}\ket{\frac{n}{4},m} &=& m\ket{\frac{n}{4},m}, \nonumber \\
    \mathcal{Y}^{2}\ket{\frac{n}{4},m} &=& \Phi \left(f_+\left(\frac{n}{4}\right)\right)\ket{\frac{n}{4},m}, \\
    \mathcal{Y}_{\pm}\ket{\frac{n}{4},m} &=& \sqrt{\Phi\left(f_+\left(\frac{n}{4}\right)\right)-  \Phi \left(f_\pm\left(m\right)\right)} \ket{\frac{n}{4},m \pm 1},\nonumber 
\end{eqnarray}
with the function $f_{\pm}(x)$ defined below Eq.~\eqref{eq:stdsu2ops2}. Note that the eigenvalue of $\mathcal{Y}_{0}$ is $n/4$, in contrast to $n/2$ for $X_0$, thus an odd $n$ cannot construct the $su(2)$ Lie algebra as $n/4$ will not be a half-integer~\cite{Bonatsos95,Gottfried03}. Using the above deformed $su(2)$ algebra the TPH Hamiltonian reads
\begin{equation}
    H_{\rm TPH} = 4Y \mathcal{Y}_{x},
\end{equation}
with $4\mathcal{Y}_{x} \coloneqq \mathcal{Y}_{+} + \mathcal{Y}_{-}$ and $4\mathcal{Y}_{y} \coloneqq -i(\mathcal{Y}_{+} - \mathcal{Y}_{-})$. The vector of operators $\mathcal{A}^{\dagger} =\left( \begin{array}{cc}
		 ba^{\dagger} &  ab^{\dagger}
		\end{array} \right)
$ can be used to uniquely represent the operators of the algebra as $4\mathcal{Y}_{\alpha} = \mathcal{A}^{\dagger}\sigma_{\alpha}\mathcal{A}$ ($\alpha = \lbrace0,x,y,z\rbrace$). This construction contrasts to the linear system that requires each element of the vector of operators $A$ to be linear in $a$ or $b$. Furthermore, it is important to note that the form of the vector of operators is not arbitrary and is linked to the underlying deformed algebra. Thus, in terms of $\mathcal{A}$, the TPH Hamiltonian reads
\begin{equation}
\begin{aligned}
   H_{\rm TPH} &= Y \mathcal{A}^{\dagger} \sigma_{x} \mathcal{A}.
\label{eq: Bogoliubov two}
\end{aligned}
\end{equation}
The above Hamiltonian can be diagonalized by using the modes $C$ and $D$,
\begin{equation}
    \left( \begin{array}{c}
		 C \\  D
		\end{array} \right) =\frac{1}{\sqrt{2}}\left( \begin{array}{cc}
		1 & 1 \\ 1 & -1
		\end{array} \right)\left( \begin{array}{c}
		ab^{\dagger} \\ ba^{\dagger}
		\end{array} \right).
\end{equation}

Note that there is no linear mapping between original field operators $(a,b)$ to $(C,D)$. Since nonlinear mode operators do not obey the standard commutation relations, i.e., $[C^{\dagger},C] = [D^{\dagger},D] = 0$, it is impossible to interpret these as `normal' modes or an excitation like in the linear case. In other words, the action of the operators $C$ ($C^{\dagger}$) and $D$ ($D^{\dagger}$) do not annihilate (create) a single excitation making it impossible to use these to evaluate the mode energy for the nonlinear Hamiltonian, as in the linear case. Due to the lack of better terminology, we will refer to these as modes throughout this paper, albeit it will be clear from the context when we refer to the nonlinear Hamiltonians. 

Therefore, to estimate the mode energy for the nonlinear Hamiltonian we begin with the eigenstate of the TPH Hamiltonian $\ket{\psi_n} \coloneqq \sum_{m} C_m\ket{\frac{n}{4},m}$, which is a linear combination of the eigenstates of $\mathcal{Y}_z$. Therefore,
\begin{eqnarray}
      H_{\rm TPH}\ket{\psi_n} &=& Y\left(\mathcal{Y}_{+}+\mathcal{Y}_{-}\right)\ket{\psi_n} \nonumber \\
      &=& E^{n}_{\rm TPH}\ket{\psi_n},
\end{eqnarray}
with
\begin{equation}
\begin{aligned}
    \label{eq:TPHenergy}
&E^{n}_{\rm TPH}\ket{\psi_n}  =  \\ &\pm\frac{1}{4}\sum_m C_m   \left[\sqrt{g_+(n,m)} + \sqrt{g_-(n,m)}\right]\ket{\frac{n}{4},m},
\end{aligned}
\end{equation}
and $g_\pm(n,m) = (n^2+4n)(n^2-4) - 16\Phi(f_\pm (m))$. The $\pm$ sign appears since the eigenvalues of $\mathcal{Y}_x$ come in pairs. The leading order of $E^{n}_{\rm TPH}$ is thus $n^2$ originating from the Casimir operator's eigenvalue. Moreover, note that even though we have $\frac{n}{2}+1$ eigenvalues $(n \in 2\mathcal{Z})$, they all follow the same leading order of $n^2$ dictated by the eigenvalue of the Casimir operator, which is independent of $m$. Inspired by the mode energy for the linear system [defined below Eq.~\eqref{eq: Bogoliubov single}] that estimates the amount of energy injected by the addition of a single-photon, we estimate the mode energy of the nonlinear Hamiltonian as \begin{equation}
\label{eq:modeenergy}
 \nu_{\rm nl} \coloneqq \frac{\partial E^{n}}{\partial n} \approxtext{$n\gg 1$} E^{n+1} - E^{n} = \nu.
\end{equation}
In the large photon limit our definition coincides with the usual definition as shown by the above equation. Thus, using the TPH energy defined in Eq.~\eqref{eq:TPHenergy}, we obtain the leading order of $\nu_{\rm TPH} \approx \omega \pm Yn$, where the first part comes from the site term $H_{\rm site}$ with the eigenvalue of $4\omega \mathcal{Y}_{0}$.  

A similar analysis can be done for the cross-phase modulation Hamiltonian $H_{\rm XPM}$. In this case, the Fock basis is the eigenbasis for the XPM Hamiltonian. The eigenvalue of $H_{\rm XPM}$ is given by,
\begin{eqnarray}
       H_{\rm XPM}|n_a,n_b\rangle &=& Za^{\dagger}a b^{\dagger} b |n_a,n_b\rangle \nonumber\\
       & =& Zn_a n_b |n_a,n_b\rangle,
\end{eqnarray}
with $E^{n}_{\rm XPM} = Zn_a n_b$. In the large photon limit, we assume that the two sites carry the equal number of photons due to inversion symmetry between sites $a$ and $b$ with $n_a \approx n_b = n/2$. Therefore, $E^{n}_{\rm XPM} \approx Zn^2/4 $ and using Eq.~\eqref{eq:modeenergy} $\nu_{\rm XPM} \approx \omega + Zn/2$, which is always positive. Note that the XPM Hamiltonian does not form a multi-dimensional subspace like TPH as the Fock state is XPM Hamiltonian's eigenstate.

In the nonlinear TPH case, we can excite the negative energy excitation mode $\nu_{\rm TPH} \approx \omega - Yn$ not only by tuning the strength $Y$ but also by increasing $n$. Whereas $\nu_{\rm XPM} \approx \omega + Z n$ is strictly positive. The above estimations we obtained coincide with the semi-classical approach outlined in Append.~\ref{sec:equation of motion}. Thus, the main difference between TPH and XPM is the possibility of a negative excitation mode $\nu_{\rm TPH} < 0$. When an external source excites the negative mode in the presence of TPH, the system can lower its energy by adding photons. In experimental setups, either other nonlinear interactions bound the system from below, or a limited number of photons are present, giving the TPH Hamiltonian a lower bound in energy.

In this section, we have investigated the quantum linear and nonlinear Hamiltonians. We have introduced the normal modes representing the energy added by a single excitation $\nu$, which corresponds to the Hamiltonian eigenvalue. To this end, we constructed the $su(2)$ algebra of the Hamiltonians. Kerr-type interaction such as XPM always gives positive excitation $\nu_{\rm XPM} = \omega + Zn$, whereas interactions that mediate hopping (non-Kerr-type) can give rise to negative excitation $\nu = \omega \pm J$ for the linear interaction and $\nu_{\rm TPH} = \omega \pm Yn$ for the TPH interaction. Moreover, we have observed that the two normal modes are completely decoupled for the linear Hamiltonian, corresponding to a uniform energy spacing. In contrast, a non-uniform spacing (i.e., a gap that depends on the number of photons $n$) appears for TPH and XPM.

\section{Transient heat transport: linear system}
\label{sec: linear thermo}

\begin{figure}
\centering
\includegraphics[width=\linewidth]{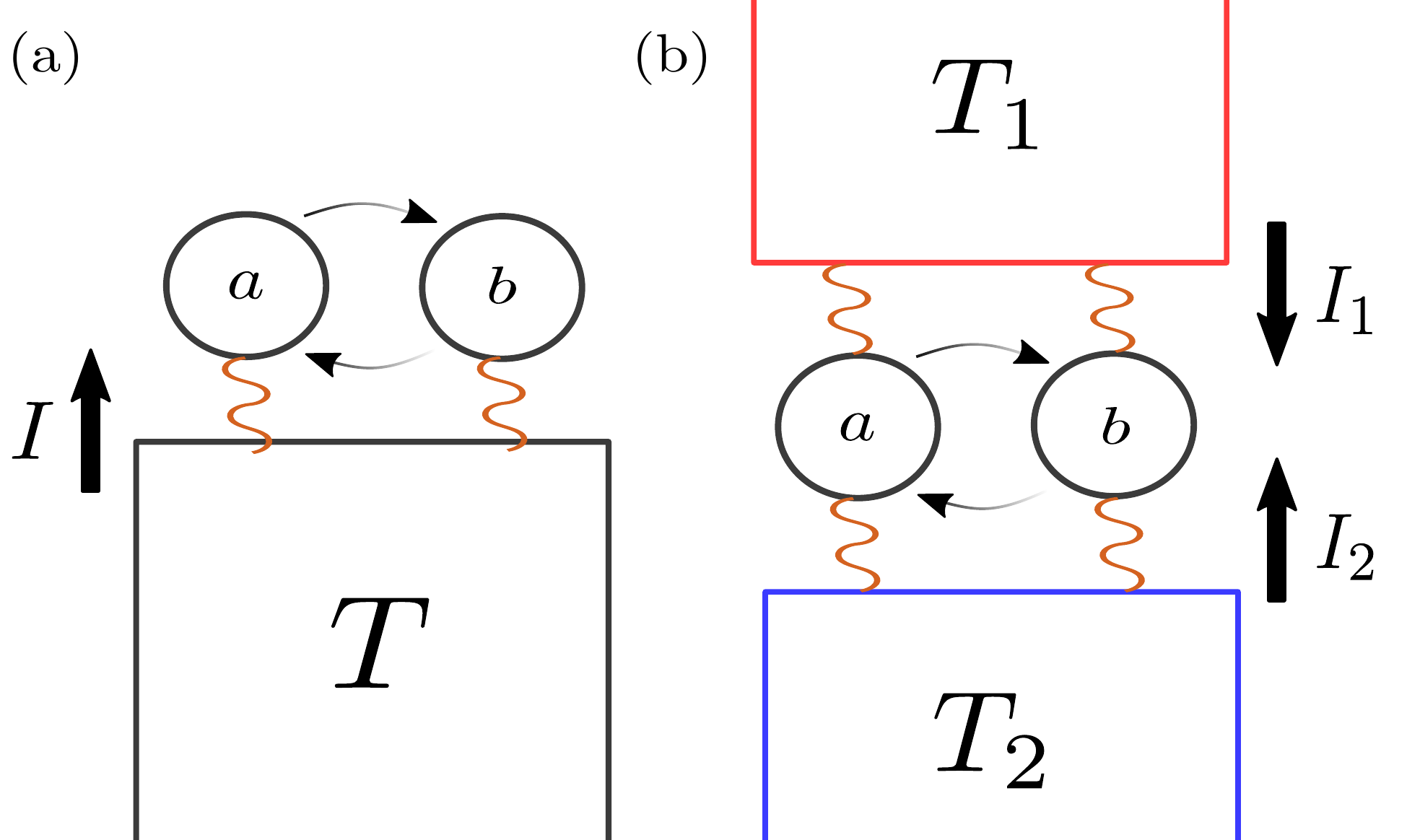}
\caption{Schematics of the models studied, considering the interaction between the dimer and (a) a single bath and (b) two baths with two different temperatures. Arrows indicate the heat current from the baths to the system, and we denote this direction as positive.}
\label{fig:model}
\end{figure}

This section builds an analytical understanding of the system's transient thermodynamics using a linear model. Namely, we consider $H = H_{\text{site}} + H_{\text{hop}}$ in the absence of nonlinear interactions. We first consider the thermodynamic response of the dimer to the single bath, as depicted in Fig.~\ref{fig:model}(a). Then, we study the two baths with different temperatures [Fig.~\ref{fig:model}(b)] to compute the nonequilibrium heat current. 

\subsection*{One bath}
We begin with the linear bosonic Hamiltonian
\begin{equation}
    H = \omega a^{\dagger}a + \omega b^{\dagger}b + J(a^{\dagger}b + b^{\dagger}a),
\label{eq:Linear hamiltonian}
\end{equation}
using the normal mode transformation described in Eq.~\eqref{eq: Bogoliubov single}, the Hamiltonian transforms into
\begin{equation}
    H = E_{+}c^{\dagger}c + E_{-}d^{\dagger}d, 
\end{equation}
where $E_{\pm} = \omega \pm J$. The corresponding eigenvalues represent the energy added by a single excitation $\nu$ and not the bare energy. 

We consider the interaction of the system with a harmonic bath described by the Hamiltonian
\begin{equation}
\begin{aligned}
    &H_{\text{B}} = \int_{0}^{\infty} d\Omega_{p} \ \Omega_{p}  \alpha_{p}^{\dagger}  \alpha_{p}
    \label{eq:bathH}
\end{aligned}
\end{equation}
with a system-bath interaction of the form
\begin{equation}
\begin{aligned}
    &H_{\text{SB}} = \int_{0}^{\infty} d\Omega_{p} \ g_{p} (a+a^{\dagger}+b+b^{\dagger}) (\alpha_{p} + \alpha_{p}^{\dagger})   \\& \qquad = \frac{1}{\sqrt{2}} \int_{0}^{\infty} d\Omega_{p} \ g_{p}(c+c^{\dagger}) \alpha_{p} +\mathrm{h.c.},
\label{eq:Hamiltonian}
\end{aligned}
\end{equation}
where $\Omega_{p} \in [0,\infty )$ is the frequency of the bath in the mode $p$ associated with the bosonic field operator at each mode $ \alpha_{p}$. The interaction strength between the system and bath $g_{p}$ is mode-dependent and gives the effective coupling strength $\Gamma \propto \int \vert g_{p} \vert^{2}$. The coupling between the external environment and each cavity is chosen such that it excites both cavities symmetrically. This choice of coupling preserves inversion symmetry, and hence the bath is only directly coupled to the symmetric mode $c$.

In order to obtain the open system dynamics, we take the standard Born-Markov secular weak-coupling approximation~\cite{Breuer02,Weiss12}. The correlation functions of the bosonic bath, $\Tr[ \alpha^{\dagger}_{p} \alpha_{q} \rho_{\text{env}}] \coloneqq \langle  \alpha^{\dagger}_{p} \alpha_{q} \rangle = \Bar{n}(\Omega_{p})\delta(\omega_{p}-\omega_{q})$, with $\Bar{n}(\Omega_{p})$ being the Bose-Einstein distributions at temperature $T$. The resultant Lindblad quantum master equation for the reduced density matrix $\rho$ is
\begin{equation}
\begin{aligned}
    \Dot{\rho}(t) =& -i[H,\rho(t)]+\Gamma E_{+} \bar{n}(E_{+}) \mathcal{D}_{ c^{\dagger}}[\rho(t)]\\
    & + \Gamma E_{+} (\bar{n}(E_{+})+1)\mathcal{D}_{ c}[\rho(t)] ,
\end{aligned}
\label{eq:linear Lindblad}
\end{equation}
where $\mathcal{D}_{ c}[\rho(t)] \coloneqq \left( 2  c\rho(t) c^{\dagger} -\lbrace  c^{\dagger} c, \rho(t)\rbrace   \right)$, with $\lbrace a,b\rbrace \coloneqq ab + ba$ being the anti-commutator. We have used a pure Ohmic spectral density function such that the electric field for each mode is $\propto \sqrt{\omega}$ in the quantized form~\cite{Breuer02,Peskin18}. We observe that the corresponding Lindblad operators are $ c$ and $ c^{\dagger}$, which indicates that the dimer problem can be reduced to the single-photon problem, as we excite only the $\nu = E_+$ mode of the system. It is important to note here that even though the problem is harmonic, most works have focused only on the steady state heat transport~\cite{Santos16} for this model.

To study transient heat currents $\Dot{Q}(t)$ we evaluate the response of the system to the thermal bath. The first law of nonequilibrium thermodynamics is, 
\begin{equation}
    \Dot{U} = \Dot{W} + \Dot{Q},
\end{equation}
where $\Dot{U},\Dot{W},\Dot{Q}$ are time derivatives of the internal energy, work done by the system, and heat supplied to the system respectively. The internal energy is $U(t) = \Tr[H\rho(t)]$ in the weak coupling regime~\cite{Alicki79} and its time derivative is
\begin{equation}
    \Dot{U} = \Tr\left[\dot{H}\rho(t)\right] + \Tr\left[H\dot{\rho}(t)\right].
\end{equation}
The first term refers to the power, and the second is the heat current~\cite{Alicki79}. As our Hamiltonian is time-independent, the energy current is equivalent to the heat current
\begin{equation}
    \Dot{Q}(t) \coloneqq I(t) \coloneqq \Tr\left[H \Dot{\rho}(t)\right].
\end{equation}
Using the dynamics from the Lindblad quantum master equation [Eq.~\eqref{eq:linear Lindblad}], we obtain the time-dependent heat current
\begin{equation}
    I(t) = 2\Gamma E_{+}^{2} \left(\bar{n}(E_{+}) - \expect{c^{\dagger} c}_t \right),
\label{eq:linear heat current1}
\end{equation}
where $\langle  c^{\dagger} c \rangle_t=\mathrm{Tr}[ c^{\dagger} c\rho(t)]$ is the expectation of the number operator in the symmetric mode. Alternately, without using the Lindblad structure the heat current can be expressed as,
\begin{eqnarray}
I(t) &=& \Tr[E_{+} c^{\dagger} c\Dot{\rho}(t)+E_{-} d^{\dagger} d\Dot{\rho}(t)] \nonumber \\
& =& E_{+}\Tr[ c^{\dagger} c\Dot{\rho}(t)] = E_{+}\frac{d}{dt}\langle  c^{\dagger} c \rangle_t.
\label{eq:HeatcurrwoLindblad}
\end{eqnarray}
Thus, equating Eqs.~\eqref{eq:linear heat current1} and~\eqref{eq:HeatcurrwoLindblad} give a linear differential equation for $\langle  c^{\dagger} c \rangle_t$ whose solution reads
\begin{equation}
    \expect{c^{\dagger} c}_t = \expect{c^{\dagger} c}_0 \ e^{-2\Gamma E_{+} t} + \bar{n}(E_{+})\left(1-e^{-2\Gamma E_{+} t}\right).
\label{eq: expectation of number}
\end{equation}
In the steady state $t\rightarrow \infty$ the above equation indicates that the expectation number in the symmetric normal mode converges to $\bar{n}(E_{+})$ as a consequence of thermalization, as expected. Plugging Eq.~\eqref{eq: expectation of number} back into Eq.~\eqref{eq:linear heat current1} yields~\cite{Xuereb15}
\begin{equation}
    I(t) = 2\Gamma E_{+}^{2} \left(\bar{n}(E_{+}) - \expect{c^{\dagger} c}_0 \right)e^{-2\Gamma E_{+} t}.
\label{eq:linear heat current2}
\end{equation}
\noindent Thus, for a system connected to a single heat bath, heat current exponentially decays with relaxation time $\tau_{R} = (2\Gamma E_{+})^{-1}$ implying that not only the decay constant $\Gamma$ but also the excitation energy of the symmetric mode plays an important role. The direction of heat current is determined by the initial expectation number of symmetric mode $ \langle  c^{\dagger} c\rangle_0$ and $\bar{n}(E_{+})$~\cite{Cuansing12}.
For example, if the initial state is chosen to be the vacuum state, the system always absorbs the bath's energy as $I(t) \geq 0, \forall t$.

\subsection*{Two baths}
Next, we study the two bath scenario as depicted in Fig.~\ref{fig:model}(b). The baths are free bosons, similar to Eq.~\eqref{eq:bathH}, with operators $ \alpha_{p}$ and $ \beta_{p}$. The interaction Hamiltonian takes the form
\begin{equation}
\begin{aligned}
    H_{\text{SB}} &= \frac{1}{\sqrt{2}}\int_{0}^{\infty} d\Omega_{p}  \  (c+c^{\dagger})\left(g_{1p} \alpha^{\dagger}_{p}+  g_{2p} \beta_{p}^{\dagger}\right)+\mathrm{h.c.}.
\label{eq:Hamiltonian2}
\end{aligned}
\end{equation}
The two baths are locally correlated, namely,
\begin{equation}
\begin{aligned}
    &\expect{  \alpha^{\dagger}_{p} \alpha_{q}} = \bar{n}_{1}(\Omega_{p})\delta(\Omega_{p}-\Omega_{q}), \\&\expect { \beta^{\dagger}_{p} \beta_{q}} = \bar{n}_{2}(\Omega_{p})\delta(\Omega_{p}-\Omega_{q}), \\&\expect{  \alpha^{\dagger}_{p} \beta_{q}} = 0,
\label{eq:correlation local bath}
\end{aligned}
\end{equation}
where $\bar{n}_{i} \ (i = 1,2)$ are Bose-Einstein distributions with temperature $T_{i}$ (the subscript $i$ refers to the bath index). In this case, the Lindblad equation reads
\begin{equation}
\begin{aligned}
    \Dot{\rho}(t) &= -i[H,\rho(t)] \\&+  E_{+}[\Gamma_{1}\bar{n}_{1}(E_{+})+\Gamma_{2}\bar{n}_{2}(E_{+})]\mathcal{D}_{ c^{\dagger}}[\rho(t)]   \\&+  E_{+}[\Gamma_{1}(\bar{n}_{1}(E_{+})+1)+\Gamma_{2}(\bar{n}_{2}(E_{+})+1)]  \mathcal{D}_{ c}[\rho(t)],
\end{aligned}
\label{eq:linear Lindblad2}
\end{equation}
where $\Gamma_{i} \ (i=1,2)$ represent effective coupling strengths to the baths. We compute the total heat current indicating the amount of heat supplied to the system, which is thus
{\small 
\begin{equation}
\begin{aligned}
&I_{\text{tot}}(t) \coloneqq I_{1}(t) + I_{2}(t) \\&= 2 E_{+}^{2}\left[\Gamma_{1}\left\{\Bar{n}_{1}(E_{+})-\expect{c^{\dagger} c}_t\right\} + \Gamma_{2}\left\{\Bar{n}_{2}(E_{+})-\expect{c^{\dagger} c}_t \right\}\right],
\label{eq:Heat current total1}
\end{aligned}
\end{equation}}
where $I_{1}$ represents the heat current from the upper bath to the system and $I_{2}$ is from the lower bath to the system. Similar to the single bath case [Eq.~\eqref{eq: expectation of number}], we obtain the expectation of the number operator of the symmetric part in time,
\begin{equation}
\begin{aligned}
    \expect{c^{\dagger} c}_t = \expect{c^{\dagger} c}_0 \ e^{-2\Gamma_{\text{T}}E_{+} t} + \Bar{n}_{\text{eff}}(E_{+})\left(1-e^{-2\Gamma_{\text{T}} E_{+} t}\right),
\label{eq: expectation of number2}
\end{aligned}
\end{equation}
where $\Gamma_{\text{T}}\Bar{n}_{\text{eff}}(E_{+}) \coloneqq \Gamma_{1}\Bar{n}_{1}(E_{+})+\Gamma_{2}\Bar{n}_{2}(E_{+})$ and $\Gamma_{\text{T}} \coloneqq \Gamma_{1} + \Gamma_{2}$. Substituting Eq.~\eqref{eq: expectation of number2} into Eq.~\eqref{eq:Heat current total1} yields
\begin{equation}
I_{\text{tot}}(t)=2 E_{+}^{2}\Gamma_{\text{T}}[\Bar{n}_{\text{eff}}(E_{+})-\expect{c^{\dagger} c}_0]e^{-2 \Gamma_{\text{T}}E_{+} t}. 
\end{equation}
If either $\Gamma_1=0$ or $\Gamma_2=0$ we obtain the previous result Eq.~\eqref{eq:linear heat current2} and our result is identical to the result in Ref.~\cite{Santos16} in the steady state limit $t\rightarrow \infty$. 

Next, we compute the net current defined as the difference of heat currents, indicating the current from the first bath to the second one. The net current reads 
\begin{equation}
\begin{aligned}
I_{\text{net}}(t) &\coloneqq  I_{1}(t) - I_{2}(t) \\& \ =  2 E_{+}^{2} \left[2\Gamma_{\text{eff}}\  \Bar{n}_{\text{diff}}(E_{+}) \right. \\& \left.+ \ \Gamma_{\text{diff}}(\Bar{n}_{\text{eff}}(E_{+})-\langle  c^{\dagger} c \rangle_0) e^{-2 \Gamma_{\text{T}}E_{+} t}\right],
\label{eq:single linear net current}
\end{aligned}
\end{equation}
where $\Gamma_{\text{eff}} \coloneqq \Gamma_{1} \Gamma_{2}/(\Gamma_{1} +\Gamma_{2}), \ \Gamma_{\text{diff}} \coloneqq \Gamma_{1} - \Gamma_{2}$, and $\Bar{n}_{\text{diff}}(E_{+}) \coloneqq \Bar{n}_{1}(E_{+}) - \Bar{n}_{2}(E_{+})$. When the system reaches the steady state, it is 
\begin{equation}
\begin{aligned}
\lim_{t\rightarrow \infty} I_{\text{net}}(t) &= 2 E_{+}^{2} \Gamma_{1}[\Bar{n}_{1}(E_{+})-\Bar{n}_{\text{eff}}(E_{+})] \\&- 2 E_{+}^{2}\Gamma_{2}[\Bar{n}_{2}(E_{+})-\Bar{n}_{\text{eff}}(E_{+})], 
\end{aligned}
\end{equation}
this result indicates that the magnitude of heat current is determined by the deviation with respect to the average number density $(\bar{n}_{\text{eff}})$. 

When the system-bath interaction strengths are identical, i.e., $\Gamma_{1} = \Gamma_{2} = \Gamma$, we obtain the celebrated Landauer-like formula~\cite{Wang08NEGF,Thingna12PRB,Wang14} from Eq.~\eqref{eq:single linear net current},
\begin{equation}
 I_{\text{net}}(t) = 2 E_{+}^{2}\Gamma [\Bar{n}_{1}(E_{+})-\Bar{n}_{2}(E_{+})].    
 \label{eq:Landauer's formula}
\end{equation}
Note that the net transient current turns out to be time-independent even though individual currents are time-dependent. The time-dependent part of Eq.~\eqref{eq:single linear net current} only depends on the $\Gamma_{\text{diff}}$, which vanishes in the presence of identical couplings. We emphasize that the Landauer-like formula has been derived only for the steady state regime, whereas here, we obtain this form even for the transient current.

In the linear response regime wherein $T_1=T+\delta T$ and $T_2=T-\delta T$, we can compute the thermal conductivity from Eq.~\eqref{eq:Landauer's formula}, which is defined by~\cite{Kane97},
\begin{equation}
    \kappa_{T} \coloneqq \frac{\partial I_{\text{net}}}{\partial T}  = \frac{16 \Gamma E_{+}^{3}}{T^{2}} \text{csch}^{2}(E_{+}/2T).
\label{eq:heat conductivity}
\end{equation}
Note that this formula holds for both transient and steady state currents. According to our assumption that $T_{1} > T_{2}$, it shows that the heat conductivity is always positive, consistent with the second law of thermodynamics. In the low-temperature limit, i.e., $E_{+}/T \gg 1$, $\kappa_{T} \propto E_{+}e^{-E_{+}/T}/T^{2}$, the conductivity goes to zero at zero temperature, and this is true in the high-temperature limit as well, i.e. $E_{+}/T \ll 1$, $\kappa_{T} \propto E_{+}$. Maximum conductivity arises when $E_{+} = 6T$.

If the baths couple to the system via the anti-symmetric mode $d$, the current can be obtained from the above expressions by replacing $c$ and $E_{+}$ by $d$ and $E_{-}$, respectively. In this case, if $E_{-} < 0$, i.e., $J > \omega$, the physical picture is similar to a bath effectively interacting with inverted oscillators. This leads to $I_{\rm net} < 0$ and $\kappa_T < 0$, which can be misconstrued as a violation of the second law or as cooling.
 
In this section, we have discussed the analytical expression of heat current in the presence of one and two baths. We have shown that the relative strength between the average number of photons of the system and the bath determines the initial heat current direction. We have computed the net current from hot to cold bath and obtained the Landauer-like formula assuring thermodynamic second law. Given the linear response regime, we analyzed thermal transport by calculating thermal conductivity. Transport is optimized at non-zero energy of the excitation mode but suppressed for both $E_{+} \rightarrow 0$ and $E_{+} \rightarrow \infty$. 

\section{Transient heat transport: nonlinear system}
\label{sec: nonlinear thermo}
In this section, we numerically evaluate the heat transport in nonlinear interactions, TPH, and XPM. We elucidate the main differences between these interactions due to the existence of {\emph {negative excitation}} mode, as we have observed in Sec.~\ref{sec:quantum nonlinear}. The analytical insight obtained using the isolated nonlinear Hamiltonian helps us understand the numerical results better and provides a qualitative analytic understanding of the main mechanisms responsible for cooling the system~\cite{Das19}.

Inspired by the linear system, we choose our system-bath interaction identical to the Eq.~\eqref{eq:Hamiltonian2}. To derive the Lindblad equation, we use the spectral decomposition of the system part of interaction Hamiltonian $H_{\text{SB}}$. The operators $c, c^{\dagger}$ are thus
\begin{equation}
    c = \sum_{n,m} c_{nm} \ket{n} \bra{m}, \quad c^{\dagger} = \sum_{n,m} c^{\ast}_{nm} \ket{m} \bra{n},
\end{equation}
where $c_{nm} \coloneqq \bra{n} c \ket{m}$ is the matrix element represented in the eigenbasis of the system Hamiltonian $\ket{n}$. Operator $c$ in the interaction picture $\tilde{c}(t) \coloneqq e^{iHt}\ c \ e^{-iHt}$ is thus
\begin{equation}
    \tilde{c}(t) \coloneqq \sum_{n,m} e^{i \Delta_{nm} t} c_{nm} \ket{n}\bra{m} \coloneqq \sum_{\nu} e^{-i \nu t} c(\nu),
\end{equation}
where $\Delta_{ij} = E_i - E_j$ and we assume that there is no degeneracy, $\Delta_{ij} \neq 0$ if $i \neq j$. The operators $c(\nu)$ ($c^{\dagger}(\nu)$) represents the annihilation (creation) operator for the specific excitation mode $\nu$ indicating  
\begin{equation}
   [H,c(\nu)] = -\nu c(\nu), \quad [H,c^{\dagger}(\nu)] = \nu c^{\dagger}(\nu),
\end{equation}
yielding the relation $c(-\nu) = c^{\dagger}(\nu)$. Therefore, the system Hamiltonian in the energy eigenbasis is
\begin{equation}
    H = \sum_{\nu} \nu c^{\dagger}(\nu)c(\nu).
\end{equation}

For the sake of simplicity, we assume that the two baths (indicated by the subscripts $1$ and $2$ below) have the same interaction strengths, i.e., $\Gamma_1 = \Gamma_2 = \Gamma$. Thus, within the Born-Markov secular and weak-coupling approximations, the Lindblad equation reads~\cite{Breuer02},
\begin{equation}
\begin{aligned}
    \dot{\rho}(t)& = -i[H,\rho(t)] \\& +\Gamma\sum_{\nu}  \nu[\bar{n}_{1}(\nu) + \bar{n}_{2}(\nu)]\mathcal{D}_{c^{\dagger}(\nu)}[\rho(t)] \\& +\Gamma\sum_{\nu}  \nu [\bar{n}_{1}(\nu) +  \bar{n}_{2}(\nu) + 2]\mathcal{D}_{c(\nu)}[\rho(t)].
\label{eq: Lindblad nonlinear}
\end{aligned}
\end{equation}
Each mode contributes to the dissipation with a rate which is a function of $\nu$. Overall, the system dissipates with a relaxation time $\tau_R = \Gamma\min{\vert\nu\vert}.$ The linear system we have considered in Sec.~\ref{sec: linear thermo} indeed exhibits two excitation modes $E_{\pm} = \omega \pm J$. Due to our specific choice of interaction, it excites only the $E_{+}$ mode [see Eq.~\eqref{eq:linear Lindblad2}]. However, this argument is not valid for a nonlinear Hamiltonian. Recall Eq.~\eqref{eq: Bogoliubov two}, the Hamiltonian in the presence of TPH is given by
\begin{equation}
    H = \omega (a^{\dagger}a + b^{\dagger}b) + Y( C^{\dagger}C - D^{\dagger}D),
\label{eq: TPH diagonalized}
\end{equation}
where
\begin{equation}
\begin{aligned}
    &C \coloneqq \frac{1}{\sqrt{2}}\left(ab^{\dagger}+ba^{\dagger}\right) = \frac{1}{\sqrt{2}}\left(c^{\dagger}c - d^{\dagger}d\right), \\&D \coloneqq \frac{1}{\sqrt{2}}(ab^{\dagger}-ba^{\dagger}) = \frac{1}{\sqrt{2}}\left(c^{\dagger}d - d^{\dagger}c\right).
\end{aligned}
\end{equation}
The basis operators $C,D$ are entangled and $[C, D] \neq 0$, composed of the single-photon basis of $c,d$. Thus, the system part of interaction Hamiltonian $c$ or $c^{\dagger}$ cannot excite a single mode but rather induces multi-mode excitations which is mode mixing. Thus, in the presence of TPH, the negative mode yielding the cooling process is inevitably excited. Once the system is in this cooling phase, the direction of heat current changes, \emph{extracting} heat from the system. It is important to note here that this cooling should not be confused with standard thermodynamic refrigeration~\cite{Levy2014}. In our case, no external work is done on the system, and cooling refers to removing the heat from the system by connecting it to a dissipative environment \cite{Zhou2015} and \emph{not} the cooling of the cold environment. This process depends on the interactions within the system and is not as universal as thermodynamic refrigeration. Moreover, our method also differs from the recently proposed sideband-like cooling that relies on an incoherent source of energy (sunlight) rather than an external work medium to cool the system~\cite{Mitchison16}. Although even in our case we require an incoherent source of energy (thermal bath) without an external work medium for cooling the system, the energy spectrum of our model that permits a negative energy mode (a condition not required in sideband-like cooling) plays a equally crucial role for the system to obtain the ground state in the cooling phase.

 In the linear system, cooling cannot occur as discussed in Sec.~\ref{sec: linear thermo} (second-last paragraph). However, in the nonlinear TPH system, the cooling can occur as a physical process when the effective interaction strength $Yn > \omega$, since the negative excitation leads to the emergence of a new ground state, as discussed below. In the linear case with the mode $E_-$ connected to the bath, the entire spectrum is inverted. In TPH, however, the spectrum is extended, comprising of both positive and negative modes and baths always exciting both the modes since they are intermingled. As we will demonstrate below via numerical simulations, the net current turns out to be always positive for all times, which supports that TPH is a physical Hamiltonian, unlike an inverted oscillator. 
  
\begin{figure}[!ht]
\centering
\includegraphics[width=\linewidth,height=7.cm]{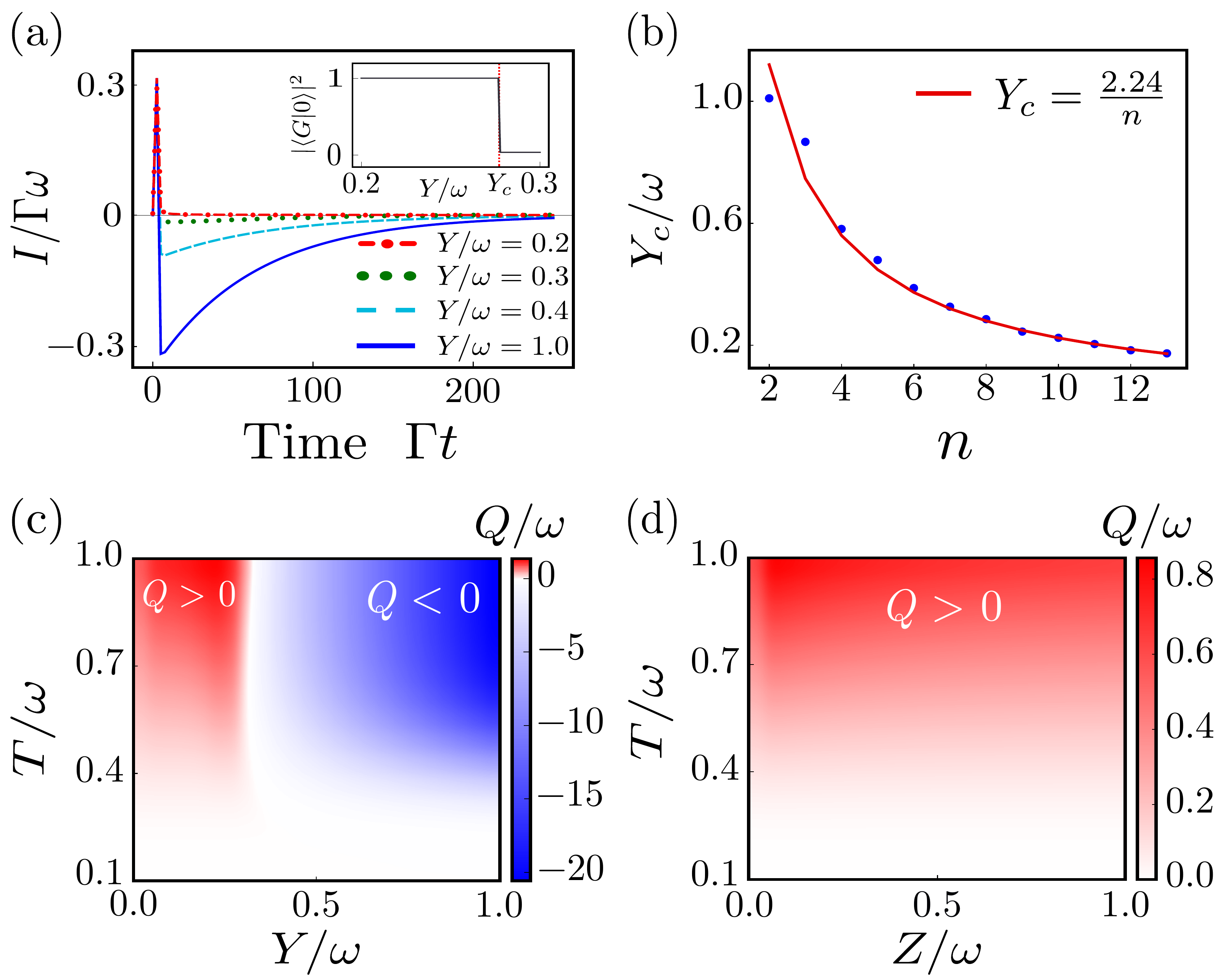}
\caption{(a) Energy current $I/\Gamma \omega$ in the presence of TPH with the temperature of the single bath $T/\omega = 0.5$. System transits from heating to cooling phase in the vicinity of $Y_{c}/\omega \approx 0.27$ where $Y_{c}$ is the critical point where the overlap between the ground state of system Hamiltonian ($\ket{G}$) and vacuum ($\ket{0}$) changes (inset). The temporal current $I/\Gamma \omega$ for $Y/\omega = 0.2$ (dashed-dotted red line) and $Y/\omega = 0.3$ (dotted green line) nearly overlap, but for $Y/\omega = 0.3 > Y_{c}/\omega$, the system starts to exhibit a negative current. (b) Scaling of the numerically evaluated critical coupling strength $Y_c/\omega$ for different values of photon number $n$ (blue closed circles). The solid red line represents the fitting function that matches well with the analytically predicted scaling of $Y_c/\omega \approx 2n^{-1}$. (c,d) Total heat supplied to the system $Q/\omega$ in the presence of TPH (c) and XPM (d) interactions as a function of the temperature of the bath and nonlinear interaction strength, where the initial state is the vacuum. The simulations use a maximum photon number of 8. The blue shaded area [right area of panel (c)] denotes the cooling regime $Q/\omega < 0$, while the red area [left area of panel (c) and full panel (d)] denotes the heating regime $Q/\omega > 0$.}
\label{fig:nonlinear current single}
\end{figure}

To verify our analytic (obtained within stringent approximations) intuition, we solve the Lindblad equation numerically by taking up to $n=8$ photons where this prudent choice of number stems from the steady state population's convergent behavior. In order to solve the Lindblad equation, with a time-independent generator, we employ the spectral-decomposition technique and numerically diagonalize the $N^2 \times N^2$ generator (with $N=\sum_{n=0}^{8}(n+1)=(n+1)(n+2)/2=45$ being the system Hilbert space dimension) obtaining all its eigenvalues and eigenvectors (left and right). We first plot the transient heat current for different strengths of TPH in the presence of a single bath with temperature $T/\omega = 0.5$ in Fig.~\ref{fig:nonlinear current single}(a). The initial state we choose is a vacuum state since it {\emph {always}} exhibits a positive current at early times, enabling us to observe a transition. Initially, regardless of the given interaction, heat is {\emph{always}} supplied to the system, as expected from the analysis of the linear system (recall that the vacuum expectation is smaller than an average number of photons of the bath). When $Y/\omega = 0.2$, the system heats up for all times. Once $Y/\omega = 0.3$, the heat current initially increases, but it becomes negative. The negative region gets larger as we increase $Y$, meaning that TPH can cool down the system. We observe that in the vicinity of the critical strength $Y_{c}/\omega \approx 0.277$ where the ground state transition occurs, the negative current starts to be induced as the negative energy gap opens up. The critical value of TPH strength $Y_c$ is when the negative excitation mode energy $\nu=\omega - Yn$ transits from being positive to negative, thus inducing the cooling process. In other words, the critical interaction strength occurs when $\nu = 0$, i.e., $Y_c/\omega = n^{-1} = 0.125$ (for $n = 8$) which is close to the numerically obtained $Y_c \approx 0.277$ in Fig.~\ref{fig:nonlinear current single}(a,c,d). The scaling of the numerically obtained $Y_c$ with $n$ is demonstrated in Fig.~\ref{fig:nonlinear current single}(b). Clearly, the actual critical parameter scales perfectly with $n^{-1}$ at large $n$. The one possible source of error in the coefficient of the analytic estimation could arise from the large $n$ limit in our analytic treatment. The critical parameter is nearly robust to temperature, as seen in Fig.~\ref{fig:nonlinear current single}(c). It is important to observe that the cooling is accompanied by a long relaxation time $\tau_{R}$ (corresponding to a metastable state), increasing with $Y$~\footnote{Qualitatively, even when we set different initial states, we obtain the negative current unless we begin with the global ground state.}. As $Y$ increases, we observe that the relaxation time $\tau_{R}$ also increases and hence for infinite TPH interaction strength $Y$ (in the cooling phase) we expect the system to always be in a metastable state for any finite time scale.

Figures~\ref{fig:nonlinear current single}(c,d) depict the total amount of energy supplied to the system, which is given by the integral of transient heat current, $Q(t) = \int_{0}^{t} dt' I(t')$. In the presence of TPH [Fig.~\ref{fig:nonlinear current single}(c)], we can extract more energy from the system as we increase both temperature and $Y$ reflecting the functional behaviour when $\nu < 0$. The system transits from a heating phase with $Y < Y_c$ to a cooling phase for $Y > Y_c$. Such a nonequilibrium phase transition does not occur in the presence of XPM [Fig.~\ref{fig:nonlinear current single}(d)] as there is no ground-state transition and also due to the positivity of the Hamiltonian (see Sec.~\ref{sec:quantum nonlinear}). An increment of $Z$ does not significantly change $Q$, and only a change in temperature affects the energy variation.  

\begin{figure}
\centering
\includegraphics[width=\linewidth,height= 10.5cm]{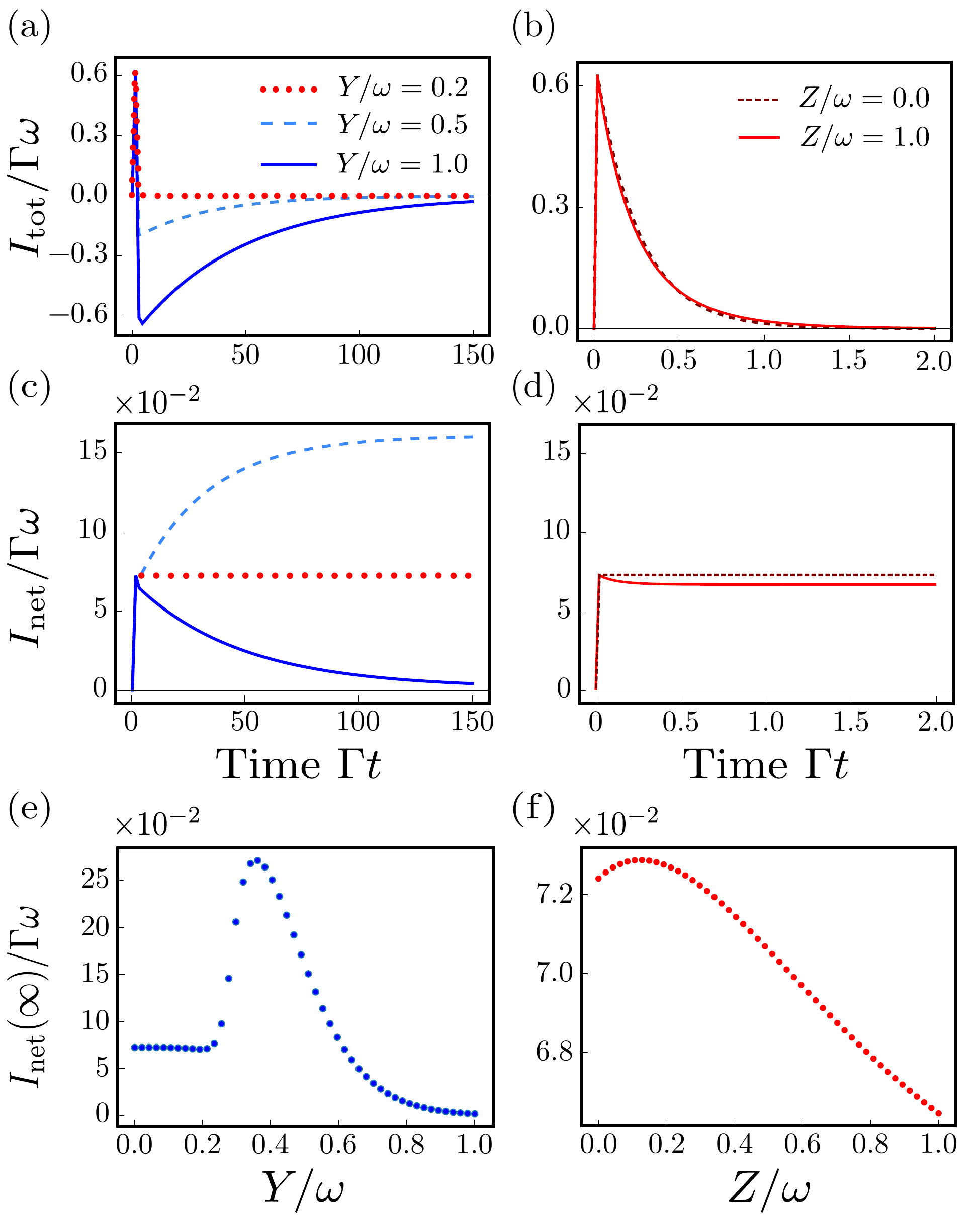}
\caption{Nonequilibrium current in the linear response regime $[(T_{1} - T_{2})/T = \Delta T = 5\%, T/\omega = 0.5, T_{1} > T_{2}]$ in the presence of TPH (first column) and XPM (second column). (a,b) Total current $I_{\text{tot}}/\Gamma \omega$ supplied to the system. Cooling is denoted by negative $I_{\text{tot}}$, whereas heating is positive. (c,d) Net current $I_{\text{net}}/\Gamma \omega$ flows from bath $1$ to $2$. In the absence of coupling, the current is time-independent, but it becomes time-dependent when interactions are present. (e,f) Nonequilibrium steady state current $I_{\rm net} (\infty)$ as a function of $Y, Z$. Peaks appear at $Y/\omega \approx 0.4$ and $Z/\omega \approx 0.2$.} 
\label{fig:noneq current}
\end{figure} 
Let us now consider the case of two baths to study the nonequilibrium transient heat current. As above, $8$ is the maximum number of photons we take into account. Fig.~\ref{fig:noneq current} illustrates the total current $I_{\text{tot}}$ in the presence of TPH (a), XPM (b) and the net current $I_{\text{net}}$ of TPH (c), XPM (d), respectively. The temperature of each bath is $T_{1/2} = T \pm \Delta T$, with $\Delta T/T = 5\%$ indicating linear response regime. The total current does not show a qualitative difference to the single bath case as the sum of individual currents simply gives it. The net current in the absence of coupling $Y = Z = 0$ corresponds to the linear system and matches the analytic result of Eq.~\eqref{eq:Landauer's formula}. The net current is no longer time-independent in the presence of nonlinear interactions (see Figs.~\ref{fig:noneq current}(c,d)), mainly due to the mixing of the normal modes $c$ and $d$. Either the net current monotonically increases ($Y/\omega = 0.5$) or is non-monotonic in time ($Y /\omega= 1$).

The nonequilibrium steady state (NESS) current is illustrated for TPH in [Fig.~\ref{fig:noneq current}(e)] and XPM in [Fig.~\ref{fig:noneq current}(f)]. When the system is in the heating phase, i.e., $Y < Y_{c}$, or in the presence of XPM, we observe almost a constant steady state current. However, once $Y > Y_{c}$, a new mode emerges, and the steady state current is enhanced up to $Y/\omega \approx 0.4$, which then decays as $Y$ increases. The decay is expected since the increase in nonlinear interaction induces stronger mixing between the modes.

\begin{figure}
\centering
\includegraphics[width=\linewidth,height= 8.0cm]{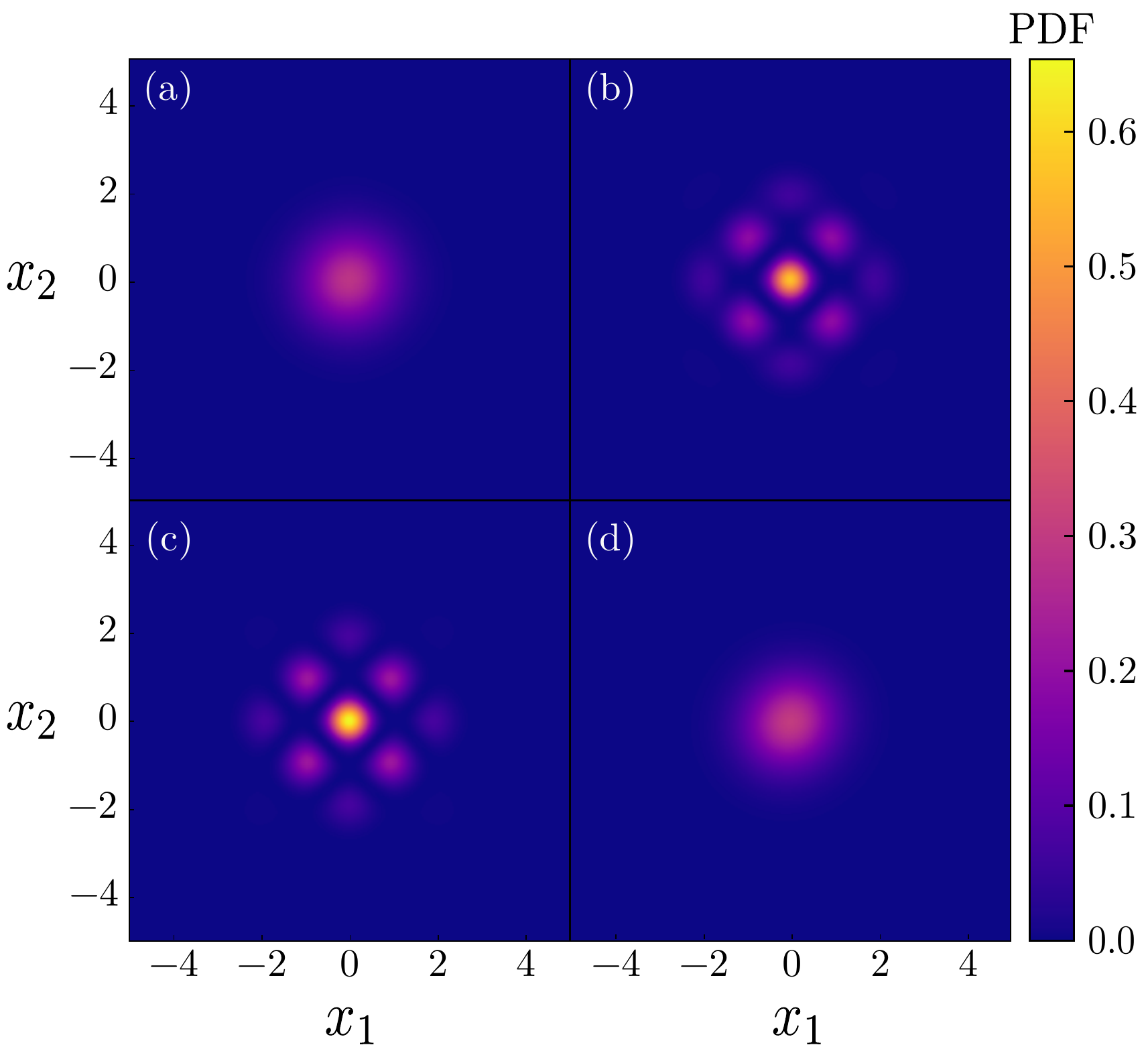}
\caption{Probability density function (PDF) of the nonequilibrium steady state $\rho_{\text{s.s.}}$ in the position representation $\langle x_{1}, x_{2} | \rho_{\text{s.s.}} | x_{1}, x_{2} \rangle$ for different nonlinear interaction strengths. (a) $Y/\omega = 0.1$, (b) $Y/\omega = 0.4$, (c) $Y/\omega = 1.0$, and (d) $Z/\omega = 1.0$. When the system experiences cooling (b,c) the above PDF exhibits a highly delocalized distribution indicating hopping between cavities.}
\label{fig:steady current position}
\end{figure}

In the heating phase [$Y < Y_{c}$, Fig.~\ref{fig:steady current position}(a) or XPM, Fig.~\ref{fig:steady current position}(d)], heat flows only through the channel associated with its cavity, i.e., the current entering into cavity $a$ escapes only through cavity $a$, as inter-cavity hopping is weak. This phenomenon is corroborated by the position representation of the steady state density matrix [Figs.~\ref{fig:steady current position}(a,d)] that shows a strong concentration at $(0,0)$, i.e., the positions of the two cavities $a$ and $b$. Once the system enters the cooling phase [$Y > Y_{c}$, Figs.~\ref{fig:steady current position}(b,c)], the hopping channel opens (as seen by the spreading of the probability density function), enabling heat flow between the two cavities. As $Y$ increases further, the hopping channel becomes dominant, causing most photons to jump between cavities. Hence, this decreases the amount of heat exchanged with the baths, enhancing the cooling effect.

In this section, we have investigated the heat transport in a nonlinear Hamiltonian in the presence of TPH and XPM, respectively. We have analyzed the cooling phase with negative excitation mode $\nu < 0$ and emphasized the differences between the linear and nonlinear Hamiltonians. In particular, we have shown that the system has both a heating and cooling phase only in the presence of TPH. 

\section{Conclusion}
\label{sec:conclusion}

In this work, we have investigated the transient and steady state thermodynamic response of a nonlinear dimer in the presence of Kerr-type cross-phase modulation (XPM) and non-Kerr-type two-photon hopping (TPH). 

We first have analyzed the linear and nonlinear Hamiltonians by constructing the relevant $su(2)$ algebra, and we have analytically predicted the appearance of a negative mode in the presence of TPH. We have then calculated the heat current analytically for the linear system. Given the vacuum initial state, heat is {\emph{always supplied}} to the system since the current direction depends on the relative strength of the system's initial expectation value and average number distribution of the bath. In the presence of TPH, the direction of transient heat current can change during the dynamics due to the existence of a negative excitation mode induced by the interaction Hamiltonian. The system thus can {\emph{emit}} heat from the system to the bath, and this phenomenon does not occur in the presence of XPM. 

For the case of two baths, we have computed net current and corresponding nonequilibrium steady state. Since the coupling between the system and each bath is identical, the transient net current becomes {\emph{time-independent}} and is given by the Landauer-like formula for the linear system. Interestingly, the nonlinear systems yield a {\emph{time-dependent}} net current that depends on the corresponding nonequilibrium steady state. Specifically, TPH allows more photons to be trapped between the two cavities (without contributing to heat transport) once the system is in the cooling phase.

Cooling quantum photonic systems due to their intrinsic interactions has not been explored before, and we expect our findings can be used to manipulate quantum states using this cooling phase. As we do not need any time modulation to control the dynamics or feedback control, this provides a novel way to achieve a coherent quantum state.

\section{Acknowledgements}
We thank Dr. Rohith Manayil for fruitful discussions. This research was supported by the Institute for Basic Science in Korea (Grant No. IBS-R024-D1 and IBS-R024-Y2). D. L. and D. G. A acknowledge support by the National Research Foundation, Prime Ministers Office, Singapore, the Ministry of Education, Singapore under the Research Centres of Excellence programme, and the Polisimulator project co-financed by Greece and the EU Regional Development Fund.

\appendix

\section{Equation of motion in semi-classical approximation}
\label{sec:equation of motion}
To reveal the difference between nonlinear interactions, let us now look at the semi-classical equations of motion. Equations when we turn off the linear coupling $(J = 0)$ read
\begin{equation}
\begin{aligned}
&i\frac{d}{dt}a_{0} = \omega a_{0} + 2Y b^{2}_{0} a^{\ast}_{0} + Z\vert b_{0}\vert^{2} a_{0}, \\& i\frac{d}{dt}b_{0} = \omega b_{0} + 2Y a^{2}_{0}b^{\ast}_{0} + Z\vert a_{0}\vert^{2}b_{0},  
\label{eq:equations of expectation values}
\end{aligned}
\end{equation}
from which it follows that
\begin{equation}
\begin{aligned}
&\frac{d}{dt}\vert a_{0} \vert^{2} = -2iY(a^{\ast 2}_{0}b^{2}_{0}- a^{2}_{0}b^{\ast 2}_{0} ), \\& \frac{d}{dt}\vert b_{0} \vert^{2} = 2iY(a^{\ast 2}_{0}b^{2}_{0}- a^{2}_{0}b^{\ast 2}_{0} ),
\label{eq:semi-classical amplitude}
\end{aligned}
\end{equation}
meaning that total power is conserved and the power at each site can change only via the TPH interaction. This is obvious once we remove one of the interactions; pure XPM $(Y = 0)$ forces the right hand side of Eq.~\eqref{eq:equations of expectation values} to be purely real, implying only oscillations of the phase. On the other hand, pure TPH $(Z = 0)$ can lead to time-dependent amplitudes since the right hand side of Eq.~\eqref{eq:equations of expectation values} becomes complex. 

\begin{figure}
\centering
\includegraphics[height = 7.5cm,width=\linewidth]{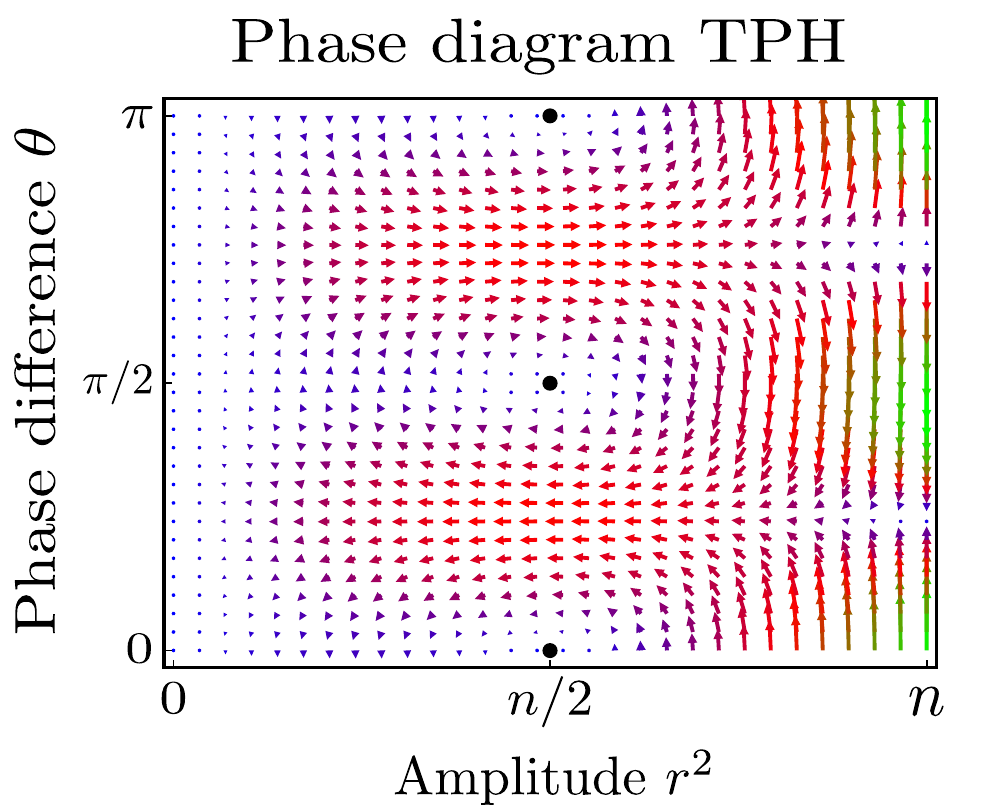}
\caption{Phase diagram of TPH for $n = 8$. Black closed circles represent the points where synchronization occurs, $r^2 = n/2$ and $\theta = 0, 0.5\pi, \pi$. The other stationary points $r = 0$, $r = \sqrt{n}$ correspond to solutions with constant amplitude.}
\label{fig:phase diagram}
\end{figure}

As the number of photons in the system is conserved, we can parameterize $a_{0} = re^{i\phi_{a}},\ b_{0} = \sqrt{n-r^{2}}e^{i\phi_{b}}$ by introducing the total number of photons $n$. Then, we obtain the equation of the phases,
\begin{equation}
\begin{aligned}
    &\frac{d\phi_{a}}{dt} = -\omega - 2Y(n-r^{2})\cos{2\theta} - Z(n-r^{2}),
    \\&\frac{d\phi_{b}}{dt} = -\omega - 2Yr^{2}\cos{2\theta} - Zr^{2},
\label{eq: Kuramoto equation}
\end{aligned}
\end{equation}
where phase difference $\theta \coloneqq  \phi_{a} - \phi_{b}$. For pure TPH, the equations above match exactly with Kuramoto equation implying synchronization and possible phase matching. Synchronization occurs when $(2r^{2}-n) = 0$ in the presence of TPH whereas for XPM we do not obtain any synchronization. As phase difference is locked by synchronization, it is possible to find the value $\theta$ from Eq.~\eqref{eq:semi-classical amplitude}, and hence the amplitude evolves according to
\begin{equation}
    r^{2}(t) =  \frac{n}{1+e^{4Yn\sin{2\theta} t}}.
\end{equation}
To obtain a stationary solution, $2r^{2} = n$, the phase difference must be $\theta = m\pi/2, m \in \mathcal{Z}$.

We now investigate the possible stationary solution of this equation and corresponding energy eigenvalue. By assuming that the system is in the stationary state, time derivative can be replaced by fixed energy value $E$. As Hamiltonian is Hermitian, $E$ should be real and this condition restricts the phase difference $\theta$ in the presence of TPH,  
\begin{equation}
2Yr(n-r^{2})\sin 2\theta = 0,
\end{equation}
$\theta = m\pi/2,\ m \in \mathcal{Z}$ or either $r = 0$ or $\sqrt{n}$. Latter condition $r = 0$ or $\sqrt{n}$ solution gives trivial fixed point where $E = \omega$. Whereas in the presence of XPM, the phase difference can be arbitrary. Non-trivial solutions of Eq.~\eqref{eq:equations of expectation values} for each case are thus
\begin{equation}
\begin{cases}
E = \omega + \frac{Z}{2} n,  \ r = \sqrt{\frac{n}{2}}, \ \theta \in [0,2\pi) & \text{(XPM)},
\\ E = \omega \pm Y n,  \ r = \sqrt{\frac{n}{2}},  \ \theta \in \lbrace 0, \frac{\pi}{2} , \pi, \frac{3\pi}{2} \rbrace & \text{(TPH)}.
\end{cases}
\end{equation}
The condition and resultant amplitude above are identical to the condition for synchronization. In other words, the stationary condition is obtained in the synchronized regime only for pure TPH. Note that even though we obtain the stationary solution of XPM, it does not imply the synchronization of cavities. Another important difference between the two interactions is a bifurcation of energy excitation mode. TPH can yield negative excitation mode as the number of photons or nonlinear interaction strength increases, while XPM exhibits only positive energy modes. 

To sum up, using the semi-classical approximation, we obtained the Kuramoto equation in the presence of pure TPH and the condition for synchronization based on Eq.~\eqref{eq: Kuramoto equation}, which is balanced amplitude between cavities with locked phase difference $m\pi/2,\ m \in \mathcal{Z}$. This coincides with the condition to reach a stationary regime. Based on the analysis above, we gain insight into the nature of the coupling Hamiltonian exhibiting inversion symmetry and the possible link of synchronizing behavior in the quantum regime to cooling~\cite{Lee2013,HerpichPRX18,Sai2020}.

\bibliographystyle{apsrev4-1}
\bibliography{Ref}

\begin{thebibliography}{62}%
\makeatletter
\providecommand \@ifxundefined [1]{%
 \@ifx{#1\undefined}
}%
\providecommand \@ifnum [1]{%
 \ifnum #1\expandafter \@firstoftwo
 \else \expandafter \@secondoftwo
 \fi
}%
\providecommand \@ifx [1]{%
 \ifx #1\expandafter \@firstoftwo
 \else \expandafter \@secondoftwo
 \fi
}%
\providecommand \natexlab [1]{#1}%
\providecommand \enquote  [1]{``#1''}%
\providecommand \bibnamefont  [1]{#1}%
\providecommand \bibfnamefont [1]{#1}%
\providecommand \citenamefont [1]{#1}%
\providecommand \href@noop [0]{\@secondoftwo}%
\providecommand \href [0]{\begingroup \@sanitize@url \@href}%
\providecommand \@href[1]{\@@startlink{#1}\@@href}%
\providecommand \@@href[1]{\endgroup#1\@@endlink}%
\providecommand \@sanitize@url [0]{\catcode `\\12\catcode `\$12\catcode
  `\&12\catcode `\#12\catcode `\^12\catcode `\_12\catcode `\%12\relax}%
\providecommand \@@startlink[1]{}%
\providecommand \@@endlink[0]{}%
\providecommand \url  [0]{\begingroup\@sanitize@url \@url }%
\providecommand \@url [1]{\endgroup\@href {#1}{\urlprefix }}%
\providecommand \urlprefix  [0]{URL }%
\providecommand \Eprint [0]{\href }%
\providecommand \doibase [0]{http://dx.doi.org/}%
\providecommand \selectlanguage [0]{\@gobble}%
\providecommand \bibinfo  [0]{\@secondoftwo}%
\providecommand \bibfield  [0]{\@secondoftwo}%
\providecommand \translation [1]{[#1]}%
\providecommand \BibitemOpen [0]{}%
\providecommand \bibitemStop [0]{}%
\providecommand \bibitemNoStop [0]{.\EOS\space}%
\providecommand \EOS [0]{\spacefactor3000\relax}%
\providecommand \BibitemShut  [1]{\csname bibitem#1\endcsname}%
\let\auto@bib@innerbib\@empty
\bibitem [{\citenamefont {Wang}\ \emph {et~al.}(2019)\citenamefont {Wang},
  \citenamefont {Sciarrino}, \citenamefont {Laing},\ and\ \citenamefont
  {Thompson}}]{Wang2019}%
  \BibitemOpen
  \bibfield  {author} {\bibinfo {author} {\bibfnamefont {J.}~\bibnamefont
  {Wang}}, \bibinfo {author} {\bibfnamefont {F.}~\bibnamefont {Sciarrino}},
  \bibinfo {author} {\bibfnamefont {A.}~\bibnamefont {Laing}}, \ and\ \bibinfo
  {author} {\bibfnamefont {M.~G.}\ \bibnamefont {Thompson}},\ }\href {\doibase
  10.1038/s41566-019-0532-1} {\bibfield  {journal} {\bibinfo  {journal} {Nat.
  Photonics}\ }\textbf {\bibinfo {volume} {14}},\ \bibinfo {pages} {273}
  (\bibinfo {year} {2019})}\BibitemShut {NoStop}%
\bibitem [{\citenamefont {Ozawa}\ \emph {et~al.}(2019)\citenamefont {Ozawa},
  \citenamefont {Price}, \citenamefont {Amo}, \citenamefont {Goldman},
  \citenamefont {Hafezi}, \citenamefont {Lu}, \citenamefont {Rechtsman},
  \citenamefont {Schuster}, \citenamefont {Simon}, \citenamefont {Zilberberg},\
  and\ \citenamefont {Carusotto}}]{Ozawa19}%
  \BibitemOpen
  \bibfield  {author} {\bibinfo {author} {\bibfnamefont {T.}~\bibnamefont
  {Ozawa}}, \bibinfo {author} {\bibfnamefont {H.~M.}\ \bibnamefont {Price}},
  \bibinfo {author} {\bibfnamefont {A.}~\bibnamefont {Amo}}, \bibinfo {author}
  {\bibfnamefont {N.}~\bibnamefont {Goldman}}, \bibinfo {author} {\bibfnamefont
  {M.}~\bibnamefont {Hafezi}}, \bibinfo {author} {\bibfnamefont
  {L.}~\bibnamefont {Lu}}, \bibinfo {author} {\bibfnamefont {M.~C.}\
  \bibnamefont {Rechtsman}}, \bibinfo {author} {\bibfnamefont {D.}~\bibnamefont
  {Schuster}}, \bibinfo {author} {\bibfnamefont {J.}~\bibnamefont {Simon}},
  \bibinfo {author} {\bibfnamefont {O.}~\bibnamefont {Zilberberg}}, \ and\
  \bibinfo {author} {\bibfnamefont {I.}~\bibnamefont {Carusotto}},\ }\href
  {\doibase 10.1103/RevModPhys.91.015006} {\bibfield  {journal} {\bibinfo
  {journal} {Rev. Mod. Phys.}\ }\textbf {\bibinfo {volume} {91}},\ \bibinfo
  {pages} {015006} (\bibinfo {year} {2019})}\BibitemShut {NoStop}%
\bibitem [{\citenamefont {Gong}\ \emph {et~al.}(2018)\citenamefont {Gong},
  \citenamefont {Ashida}, \citenamefont {Kawabata}, \citenamefont {Takasan},
  \citenamefont {Higashikawa},\ and\ \citenamefont {Ueda}}]{Gong18}%
  \BibitemOpen
  \bibfield  {author} {\bibinfo {author} {\bibfnamefont {Z.}~\bibnamefont
  {Gong}}, \bibinfo {author} {\bibfnamefont {Y.}~\bibnamefont {Ashida}},
  \bibinfo {author} {\bibfnamefont {K.}~\bibnamefont {Kawabata}}, \bibinfo
  {author} {\bibfnamefont {K.}~\bibnamefont {Takasan}}, \bibinfo {author}
  {\bibfnamefont {S.}~\bibnamefont {Higashikawa}}, \ and\ \bibinfo {author}
  {\bibfnamefont {M.}~\bibnamefont {Ueda}},\ }\href {\doibase
  10.1103/PhysRevX.8.031079} {\bibfield  {journal} {\bibinfo  {journal} {Phys.
  Rev. X}\ }\textbf {\bibinfo {volume} {8}},\ \bibinfo {pages} {031079}
  (\bibinfo {year} {2018})}\BibitemShut {NoStop}%
\bibitem [{\citenamefont {Harari}\ \emph {et~al.}(2018)\citenamefont {Harari},
  \citenamefont {Bandres}, \citenamefont {Lumer}, \citenamefont {Rechtsman},
  \citenamefont {Chong}, \citenamefont {Khajavikhan}, \citenamefont
  {Christodoulides},\ and\ \citenamefont {Segev}}]{Harari18}%
  \BibitemOpen
  \bibfield  {author} {\bibinfo {author} {\bibfnamefont {G.}~\bibnamefont
  {Harari}}, \bibinfo {author} {\bibfnamefont {M.~A.}\ \bibnamefont {Bandres}},
  \bibinfo {author} {\bibfnamefont {Y.}~\bibnamefont {Lumer}}, \bibinfo
  {author} {\bibfnamefont {M.~C.}\ \bibnamefont {Rechtsman}}, \bibinfo {author}
  {\bibfnamefont {Y.~D.}\ \bibnamefont {Chong}}, \bibinfo {author}
  {\bibfnamefont {M.}~\bibnamefont {Khajavikhan}}, \bibinfo {author}
  {\bibfnamefont {D.~N.}\ \bibnamefont {Christodoulides}}, \ and\ \bibinfo
  {author} {\bibfnamefont {M.}~\bibnamefont {Segev}},\ }\href {\doibase
  10.1126/science.aar4003} {\bibfield  {journal} {\bibinfo  {journal}
  {Science}\ }\textbf {\bibinfo {volume} {359}},\ \bibinfo {pages} {eaar4003}
  (\bibinfo {year} {2018})}\BibitemShut {NoStop}%
\bibitem [{\citenamefont {Bandres}\ \emph {et~al.}(2018)\citenamefont
  {Bandres}, \citenamefont {Wittek}, \citenamefont {Harari}, \citenamefont
  {Parto}, \citenamefont {Ren}, \citenamefont {Segev}, \citenamefont
  {Christodoulides},\ and\ \citenamefont {Khajavikhan}}]{Miguel18}%
  \BibitemOpen
  \bibfield  {author} {\bibinfo {author} {\bibfnamefont {M.~A.}\ \bibnamefont
  {Bandres}}, \bibinfo {author} {\bibfnamefont {S.}~\bibnamefont {Wittek}},
  \bibinfo {author} {\bibfnamefont {G.}~\bibnamefont {Harari}}, \bibinfo
  {author} {\bibfnamefont {M.}~\bibnamefont {Parto}}, \bibinfo {author}
  {\bibfnamefont {J.}~\bibnamefont {Ren}}, \bibinfo {author} {\bibfnamefont
  {M.}~\bibnamefont {Segev}}, \bibinfo {author} {\bibfnamefont {D.~N.}\
  \bibnamefont {Christodoulides}}, \ and\ \bibinfo {author} {\bibfnamefont
  {M.}~\bibnamefont {Khajavikhan}},\ }\href {\doibase 10.1126/science.aar4005}
  {\bibfield  {journal} {\bibinfo  {journal} {Science}\ }\textbf {\bibinfo
  {volume} {359}},\ \bibinfo {pages} {eaar4005} (\bibinfo {year}
  {2018})}\BibitemShut {NoStop}%
\bibitem [{\citenamefont {Rechtsman}\ \emph {et~al.}(2016)\citenamefont
  {Rechtsman}, \citenamefont {Lumer}, \citenamefont {Plotnik}, \citenamefont
  {Perez-Leija}, \citenamefont {Szameit},\ and\ \citenamefont
  {Segev}}]{Rechtsman16}%
  \BibitemOpen
  \bibfield  {author} {\bibinfo {author} {\bibfnamefont {M.~C.}\ \bibnamefont
  {Rechtsman}}, \bibinfo {author} {\bibfnamefont {Y.}~\bibnamefont {Lumer}},
  \bibinfo {author} {\bibfnamefont {Y.}~\bibnamefont {Plotnik}}, \bibinfo
  {author} {\bibfnamefont {A.}~\bibnamefont {Perez-Leija}}, \bibinfo {author}
  {\bibfnamefont {A.}~\bibnamefont {Szameit}}, \ and\ \bibinfo {author}
  {\bibfnamefont {M.}~\bibnamefont {Segev}},\ }\href {\doibase
  10.1364/OPTICA.3.000925} {\bibfield  {journal} {\bibinfo  {journal} {Optica}\
  }\textbf {\bibinfo {volume} {3}},\ \bibinfo {pages} {925} (\bibinfo {year}
  {2016})}\BibitemShut {NoStop}%
\bibitem [{\citenamefont {Mittal}\ \emph {et~al.}(2016)\citenamefont {Mittal},
  \citenamefont {Orre},\ and\ \citenamefont {Hafezi}}]{Mittal16E}%
  \BibitemOpen
  \bibfield  {author} {\bibinfo {author} {\bibfnamefont {S.}~\bibnamefont
  {Mittal}}, \bibinfo {author} {\bibfnamefont {V.~V.}\ \bibnamefont {Orre}}, \
  and\ \bibinfo {author} {\bibfnamefont {M.}~\bibnamefont {Hafezi}},\ }\href
  {\doibase 10.1364/OE.24.015631} {\bibfield  {journal} {\bibinfo  {journal}
  {Opt. Express}\ }\textbf {\bibinfo {volume} {24}},\ \bibinfo {pages} {15631}
  (\bibinfo {year} {2016})}\BibitemShut {NoStop}%
\bibitem [{\citenamefont {Tambasco}\ \emph {et~al.}(2018)\citenamefont
  {Tambasco}, \citenamefont {Corrielli}, \citenamefont {Chapman}, \citenamefont
  {Crespi}, \citenamefont {Zilberberg}, \citenamefont {Osellame},\ and\
  \citenamefont {Peruzzo}}]{Tambasco18}%
  \BibitemOpen
  \bibfield  {author} {\bibinfo {author} {\bibfnamefont {J.-L.}\ \bibnamefont
  {Tambasco}}, \bibinfo {author} {\bibfnamefont {G.}~\bibnamefont {Corrielli}},
  \bibinfo {author} {\bibfnamefont {R.~J.}\ \bibnamefont {Chapman}}, \bibinfo
  {author} {\bibfnamefont {A.}~\bibnamefont {Crespi}}, \bibinfo {author}
  {\bibfnamefont {O.}~\bibnamefont {Zilberberg}}, \bibinfo {author}
  {\bibfnamefont {R.}~\bibnamefont {Osellame}}, \ and\ \bibinfo {author}
  {\bibfnamefont {A.}~\bibnamefont {Peruzzo}},\ }\href {\doibase
  10.1126/sciadv.aat3187} {\bibfield  {journal} {\bibinfo  {journal} {Sci.
  Adv.}\ }\textbf {\bibinfo {volume} {4}},\ \bibinfo {pages} {eaat3187}
  (\bibinfo {year} {2018})}\BibitemShut {NoStop}%
\bibitem [{\citenamefont {Gneiting}\ \emph {et~al.}(2019)\citenamefont
  {Gneiting}, \citenamefont {Leykam},\ and\ \citenamefont {Nori}}]{Clemens19}%
  \BibitemOpen
  \bibfield  {author} {\bibinfo {author} {\bibfnamefont {C.}~\bibnamefont
  {Gneiting}}, \bibinfo {author} {\bibfnamefont {D.}~\bibnamefont {Leykam}}, \
  and\ \bibinfo {author} {\bibfnamefont {F.}~\bibnamefont {Nori}},\ }\href
  {\doibase 10.1103/PhysRevLett.122.066601} {\bibfield  {journal} {\bibinfo
  {journal} {Phys. Rev. Lett.}\ }\textbf {\bibinfo {volume} {122}},\ \bibinfo
  {pages} {066601} (\bibinfo {year} {2019})}\BibitemShut {NoStop}%
\bibitem [{\citenamefont {Yuan}\ \emph {et~al.}(2019)\citenamefont {Yuan},
  \citenamefont {Lin}, \citenamefont {Zhang}, \citenamefont {Xiao},
  \citenamefont {Chen},\ and\ \citenamefont {Fan}}]{Yuan19}%
  \BibitemOpen
  \bibfield  {author} {\bibinfo {author} {\bibfnamefont {L.}~\bibnamefont
  {Yuan}}, \bibinfo {author} {\bibfnamefont {Q.}~\bibnamefont {Lin}}, \bibinfo
  {author} {\bibfnamefont {A.}~\bibnamefont {Zhang}}, \bibinfo {author}
  {\bibfnamefont {M.}~\bibnamefont {Xiao}}, \bibinfo {author} {\bibfnamefont
  {X.}~\bibnamefont {Chen}}, \ and\ \bibinfo {author} {\bibfnamefont
  {S.}~\bibnamefont {Fan}},\ }\href {\doibase 10.1103/PhysRevLett.122.083903}
  {\bibfield  {journal} {\bibinfo  {journal} {Phys. Rev. Lett.}\ }\textbf
  {\bibinfo {volume} {122}},\ \bibinfo {pages} {083903} (\bibinfo {year}
  {2019})}\BibitemShut {NoStop}%
\bibitem [{\citenamefont {Han}\ \emph {et~al.}(2020)\citenamefont {Han},
  \citenamefont {Sukhorukov},\ and\ \citenamefont {Leykam}}]{Han20}%
  \BibitemOpen
  \bibfield  {author} {\bibinfo {author} {\bibfnamefont {J.}~\bibnamefont
  {Han}}, \bibinfo {author} {\bibfnamefont {A.~A.}\ \bibnamefont {Sukhorukov}},
  \ and\ \bibinfo {author} {\bibfnamefont {D.}~\bibnamefont {Leykam}},\ }\href
  {\doibase 10.1364/PRJ.399919} {\bibfield  {journal} {\bibinfo  {journal}
  {Photon. Res.}\ }\textbf {\bibinfo {volume} {8}},\ \bibinfo {pages} {B15}
  (\bibinfo {year} {2020})}\BibitemShut {NoStop}%
\bibitem [{\citenamefont {Chang}\ \emph {et~al.}(2014)\citenamefont {Chang},
  \citenamefont {Vuletic},\ and\ \citenamefont {Lukin}}]{Chang14}%
  \BibitemOpen
  \bibfield  {author} {\bibinfo {author} {\bibfnamefont {D.~E.}\ \bibnamefont
  {Chang}}, \bibinfo {author} {\bibfnamefont {V.}~\bibnamefont {Vuletic}}, \
  and\ \bibinfo {author} {\bibfnamefont {M.~D.}\ \bibnamefont {Lukin}},\ }\href
  {\doibase 10.1038/nphoton.2014.192} {\bibfield  {journal} {\bibinfo
  {journal} {Nat. Photonics}\ }\textbf {\bibinfo {volume} {8}},\ \bibinfo
  {pages} {685} (\bibinfo {year} {2014})}\BibitemShut {NoStop}%
\bibitem [{\citenamefont {Noh}\ and\ \citenamefont
  {Angelakis}(2016)}]{Noh_2016}%
  \BibitemOpen
  \bibfield  {author} {\bibinfo {author} {\bibfnamefont {C.}~\bibnamefont
  {Noh}}\ and\ \bibinfo {author} {\bibfnamefont {D.~G.}\ \bibnamefont
  {Angelakis}},\ }\href {\doibase 10.1088/0034-4885/80/1/016401} {\bibfield
  {journal} {\bibinfo  {journal} {Rep. Prog. Phys.}\ }\textbf {\bibinfo
  {volume} {80}},\ \bibinfo {pages} {016401} (\bibinfo {year}
  {2016})}\BibitemShut {NoStop}%
\bibitem [{\citenamefont {Smirnova}\ \emph {et~al.}(2020)\citenamefont
  {Smirnova}, \citenamefont {Leykam}, \citenamefont {Chong},\ and\
  \citenamefont {Kivshar}}]{Smirnova20}%
  \BibitemOpen
  \bibfield  {author} {\bibinfo {author} {\bibfnamefont {D.}~\bibnamefont
  {Smirnova}}, \bibinfo {author} {\bibfnamefont {D.}~\bibnamefont {Leykam}},
  \bibinfo {author} {\bibfnamefont {Y.}~\bibnamefont {Chong}}, \ and\ \bibinfo
  {author} {\bibfnamefont {Y.}~\bibnamefont {Kivshar}},\ }\href {\doibase
  10.1063/1.5142397} {\bibfield  {journal} {\bibinfo  {journal} {Appl. Phys.
  Rev.}\ }\textbf {\bibinfo {volume} {7}},\ \bibinfo {pages} {021306} (\bibinfo
  {year} {2020})}\BibitemShut {NoStop}%
\bibitem [{\citenamefont {Roushan}\ \emph {et~al.}(2017)\citenamefont
  {Roushan}, \citenamefont {Neill}, \citenamefont {Tangpanitanon},
  \citenamefont {Bastidas}, \citenamefont {Megrant}, \citenamefont {Barends},
  \citenamefont {Chen}, \citenamefont {Chen}, \citenamefont {Chiaro},
  \citenamefont {Dunsworth}, \citenamefont {Fowler}, \citenamefont {Foxen},
  \citenamefont {Giustina}, \citenamefont {Jeffrey}, \citenamefont {Kelly},
  \citenamefont {Lucero}, \citenamefont {Mutus}, \citenamefont {Neeley},
  \citenamefont {Quintana}, \citenamefont {Sank}, \citenamefont {Vainsencher},
  \citenamefont {Wenner}, \citenamefont {White}, \citenamefont {Neven},
  \citenamefont {Angelakis},\ and\ \citenamefont {Martinis}}]{Roushan1175}%
  \BibitemOpen
  \bibfield  {author} {\bibinfo {author} {\bibfnamefont {P.}~\bibnamefont
  {Roushan}}, \bibinfo {author} {\bibfnamefont {C.}~\bibnamefont {Neill}},
  \bibinfo {author} {\bibfnamefont {J.}~\bibnamefont {Tangpanitanon}}, \bibinfo
  {author} {\bibfnamefont {V.~M.}\ \bibnamefont {Bastidas}}, \bibinfo {author}
  {\bibfnamefont {A.}~\bibnamefont {Megrant}}, \bibinfo {author} {\bibfnamefont
  {R.}~\bibnamefont {Barends}}, \bibinfo {author} {\bibfnamefont
  {Y.}~\bibnamefont {Chen}}, \bibinfo {author} {\bibfnamefont {Z.}~\bibnamefont
  {Chen}}, \bibinfo {author} {\bibfnamefont {B.}~\bibnamefont {Chiaro}},
  \bibinfo {author} {\bibfnamefont {A.}~\bibnamefont {Dunsworth}}, \bibinfo
  {author} {\bibfnamefont {A.}~\bibnamefont {Fowler}}, \bibinfo {author}
  {\bibfnamefont {B.}~\bibnamefont {Foxen}}, \bibinfo {author} {\bibfnamefont
  {M.}~\bibnamefont {Giustina}}, \bibinfo {author} {\bibfnamefont
  {E.}~\bibnamefont {Jeffrey}}, \bibinfo {author} {\bibfnamefont
  {J.}~\bibnamefont {Kelly}}, \bibinfo {author} {\bibfnamefont
  {E.}~\bibnamefont {Lucero}}, \bibinfo {author} {\bibfnamefont
  {J.}~\bibnamefont {Mutus}}, \bibinfo {author} {\bibfnamefont
  {M.}~\bibnamefont {Neeley}}, \bibinfo {author} {\bibfnamefont
  {C.}~\bibnamefont {Quintana}}, \bibinfo {author} {\bibfnamefont
  {D.}~\bibnamefont {Sank}}, \bibinfo {author} {\bibfnamefont {A.}~\bibnamefont
  {Vainsencher}}, \bibinfo {author} {\bibfnamefont {J.}~\bibnamefont {Wenner}},
  \bibinfo {author} {\bibfnamefont {T.}~\bibnamefont {White}}, \bibinfo
  {author} {\bibfnamefont {H.}~\bibnamefont {Neven}}, \bibinfo {author}
  {\bibfnamefont {D.~G.}\ \bibnamefont {Angelakis}}, \ and\ \bibinfo {author}
  {\bibfnamefont {J.}~\bibnamefont {Martinis}},\ }\href {\doibase
  10.1126/science.aao1401} {\bibfield  {journal} {\bibinfo  {journal}
  {Science}\ }\textbf {\bibinfo {volume} {358}},\ \bibinfo {pages} {1175}
  (\bibinfo {year} {2017})}\BibitemShut {NoStop}%
\bibitem [{\citenamefont {Blais}\ \emph {et~al.}(2021)\citenamefont {Blais},
  \citenamefont {Grimsmo}, \citenamefont {Girvin},\ and\ \citenamefont
  {Wallraff}}]{Blais20}%
  \BibitemOpen
  \bibfield  {author} {\bibinfo {author} {\bibfnamefont {A.}~\bibnamefont
  {Blais}}, \bibinfo {author} {\bibfnamefont {A.~L.}\ \bibnamefont {Grimsmo}},
  \bibinfo {author} {\bibfnamefont {S.~M.}\ \bibnamefont {Girvin}}, \ and\
  \bibinfo {author} {\bibfnamefont {A.}~\bibnamefont {Wallraff}},\ }\href
  {\doibase 10.1103/RevModPhys.93.025005} {\bibfield  {journal} {\bibinfo
  {journal} {Rev. Mod. Phys.}\ }\textbf {\bibinfo {volume} {93}},\ \bibinfo
  {pages} {025005} (\bibinfo {year} {2021})}\BibitemShut {NoStop}%
\bibitem [{\citenamefont {Altland}\ and\ \citenamefont
  {Simons}(2010)}]{altland10}%
  \BibitemOpen
  \bibfield  {author} {\bibinfo {author} {\bibfnamefont {A.}~\bibnamefont
  {Altland}}\ and\ \bibinfo {author} {\bibfnamefont {B.~D.}\ \bibnamefont
  {Simons}},\ }\href@noop {} {\emph {\bibinfo {title} {Condensed matter field
  theory}}}\ (\bibinfo  {publisher} {Cambridge university press},\ \bibinfo
  {year} {2010})\BibitemShut {NoStop}%
\bibitem [{\citenamefont {Xu}\ \emph {et~al.}(2017)\citenamefont {Xu},
  \citenamefont {Thingna},\ and\ \citenamefont {Wang}}]{XuPRB17}%
  \BibitemOpen
  \bibfield  {author} {\bibinfo {author} {\bibfnamefont {X.}~\bibnamefont
  {Xu}}, \bibinfo {author} {\bibfnamefont {J.}~\bibnamefont {Thingna}}, \ and\
  \bibinfo {author} {\bibfnamefont {J.-S.}\ \bibnamefont {Wang}},\ }\href
  {\doibase 10.1103/PhysRevB.95.035428} {\bibfield  {journal} {\bibinfo
  {journal} {Phys. Rev. B}\ }\textbf {\bibinfo {volume} {95}},\ \bibinfo
  {pages} {035428} (\bibinfo {year} {2017})}\BibitemShut {NoStop}%
\bibitem [{\citenamefont {Greiner}\ \emph
  {et~al.}(2002{\natexlab{a}})\citenamefont {Greiner}, \citenamefont {Mandel},
  \citenamefont {Esslinger}, \citenamefont {H{\"a}nsch},\ and\ \citenamefont
  {Bloch}}]{Greiner02PT}%
  \BibitemOpen
  \bibfield  {author} {\bibinfo {author} {\bibfnamefont {M.}~\bibnamefont
  {Greiner}}, \bibinfo {author} {\bibfnamefont {O.}~\bibnamefont {Mandel}},
  \bibinfo {author} {\bibfnamefont {T.}~\bibnamefont {Esslinger}}, \bibinfo
  {author} {\bibfnamefont {T.~W.}\ \bibnamefont {H{\"a}nsch}}, \ and\ \bibinfo
  {author} {\bibfnamefont {I.}~\bibnamefont {Bloch}},\ }\href {\doibase
  10.1038/415039a} {\bibfield  {journal} {\bibinfo  {journal} {Nature}\
  }\textbf {\bibinfo {volume} {415}},\ \bibinfo {pages} {39} (\bibinfo {year}
  {2002}{\natexlab{a}})}\BibitemShut {NoStop}%
\bibitem [{\citenamefont {Greiner}\ \emph
  {et~al.}(2002{\natexlab{b}})\citenamefont {Greiner}, \citenamefont {Mandel},
  \citenamefont {H{\"a}nsch},\ and\ \citenamefont {Bloch}}]{Greiner02}%
  \BibitemOpen
  \bibfield  {author} {\bibinfo {author} {\bibfnamefont {M.}~\bibnamefont
  {Greiner}}, \bibinfo {author} {\bibfnamefont {O.}~\bibnamefont {Mandel}},
  \bibinfo {author} {\bibfnamefont {T.~W.}\ \bibnamefont {H{\"a}nsch}}, \ and\
  \bibinfo {author} {\bibfnamefont {I.}~\bibnamefont {Bloch}},\ }\href
  {\doibase 10.1038/nature00968} {\bibfield  {journal} {\bibinfo  {journal}
  {Nature}\ }\textbf {\bibinfo {volume} {419}},\ \bibinfo {pages} {51}
  (\bibinfo {year} {2002}{\natexlab{b}})}\BibitemShut {NoStop}%
\bibitem [{\citenamefont {Stepanenko}\ and\ \citenamefont
  {Gorlach}(2020)}]{StepanenkoPRA20}%
  \BibitemOpen
  \bibfield  {author} {\bibinfo {author} {\bibfnamefont {A.~A.}\ \bibnamefont
  {Stepanenko}}\ and\ \bibinfo {author} {\bibfnamefont {M.~A.}\ \bibnamefont
  {Gorlach}},\ }\href {\doibase 10.1103/PhysRevA.102.013510} {\bibfield
  {journal} {\bibinfo  {journal} {Phys. Rev. A}\ }\textbf {\bibinfo {volume}
  {102}},\ \bibinfo {pages} {013510} (\bibinfo {year} {2020})}\BibitemShut
  {NoStop}%
\bibitem [{\citenamefont {Razzari}\ \emph {et~al.}(2010)\citenamefont
  {Razzari}, \citenamefont {Duchesne}, \citenamefont {Ferrera}, \citenamefont
  {Morandotti}, \citenamefont {Chu}, \citenamefont {Little},\ and\
  \citenamefont {Moss}}]{Razzari2010}%
  \BibitemOpen
  \bibfield  {author} {\bibinfo {author} {\bibfnamefont {L.}~\bibnamefont
  {Razzari}}, \bibinfo {author} {\bibfnamefont {D.}~\bibnamefont {Duchesne}},
  \bibinfo {author} {\bibfnamefont {M.}~\bibnamefont {Ferrera}}, \bibinfo
  {author} {\bibfnamefont {R.}~\bibnamefont {Morandotti}}, \bibinfo {author}
  {\bibfnamefont {S.}~\bibnamefont {Chu}}, \bibinfo {author} {\bibfnamefont
  {B.~E.}\ \bibnamefont {Little}}, \ and\ \bibinfo {author} {\bibfnamefont
  {D.~J.}\ \bibnamefont {Moss}},\ }\href {\doibase 10.1038/nphoton.2009.236}
  {\bibfield  {journal} {\bibinfo  {journal} {Nat. Photonics}\ }\textbf
  {\bibinfo {volume} {4}},\ \bibinfo {pages} {41} (\bibinfo {year}
  {2010})}\BibitemShut {NoStop}%
\bibitem [{\citenamefont {Krachmalnicoff}\ \emph {et~al.}(2010)\citenamefont
  {Krachmalnicoff}, \citenamefont {Jaskula}, \citenamefont {Bonneau},
  \citenamefont {Leung}, \citenamefont {Partridge}, \citenamefont {Boiron},
  \citenamefont {Westbrook}, \citenamefont {Deuar}, \citenamefont
  {Zi\ifmmode~\acute{n}\else \'{n}\fi{}}, \citenamefont {Trippenbach},\ and\
  \citenamefont {Kheruntsyan}}]{Krachmalnicoff10}%
  \BibitemOpen
  \bibfield  {author} {\bibinfo {author} {\bibfnamefont {V.}~\bibnamefont
  {Krachmalnicoff}}, \bibinfo {author} {\bibfnamefont {J.-C.}\ \bibnamefont
  {Jaskula}}, \bibinfo {author} {\bibfnamefont {M.}~\bibnamefont {Bonneau}},
  \bibinfo {author} {\bibfnamefont {V.}~\bibnamefont {Leung}}, \bibinfo
  {author} {\bibfnamefont {G.~B.}\ \bibnamefont {Partridge}}, \bibinfo {author}
  {\bibfnamefont {D.}~\bibnamefont {Boiron}}, \bibinfo {author} {\bibfnamefont
  {C.~I.}\ \bibnamefont {Westbrook}}, \bibinfo {author} {\bibfnamefont
  {P.}~\bibnamefont {Deuar}}, \bibinfo {author} {\bibfnamefont
  {P.}~\bibnamefont {Zi\ifmmode~\acute{n}\else \'{n}\fi{}}}, \bibinfo {author}
  {\bibfnamefont {M.}~\bibnamefont {Trippenbach}}, \ and\ \bibinfo {author}
  {\bibfnamefont {K.~V.}\ \bibnamefont {Kheruntsyan}},\ }\href {\doibase
  10.1103/PhysRevLett.104.150402} {\bibfield  {journal} {\bibinfo  {journal}
  {Phys. Rev. Lett.}\ }\textbf {\bibinfo {volume} {104}},\ \bibinfo {pages}
  {150402} (\bibinfo {year} {2010})}\BibitemShut {NoStop}%
\bibitem [{\citenamefont {Ding}\ \emph
  {et~al.}(2017{\natexlab{a}})\citenamefont {Ding}, \citenamefont
  {Maslennikov}, \citenamefont {Habl\"utzel},\ and\ \citenamefont
  {Matsukevich}}]{Ding2017}%
  \BibitemOpen
  \bibfield  {author} {\bibinfo {author} {\bibfnamefont {S.}~\bibnamefont
  {Ding}}, \bibinfo {author} {\bibfnamefont {G.}~\bibnamefont {Maslennikov}},
  \bibinfo {author} {\bibfnamefont {R.}~\bibnamefont {Habl\"utzel}}, \ and\
  \bibinfo {author} {\bibfnamefont {D.}~\bibnamefont {Matsukevich}},\ }\href
  {\doibase 10.1103/PhysRevLett.119.193602} {\bibfield  {journal} {\bibinfo
  {journal} {Phys. Rev. Lett.}\ }\textbf {\bibinfo {volume} {119}},\ \bibinfo
  {pages} {193602} (\bibinfo {year} {2017}{\natexlab{a}})}\BibitemShut
  {NoStop}%
\bibitem [{\citenamefont {Ding}\ \emph
  {et~al.}(2017{\natexlab{b}})\citenamefont {Ding}, \citenamefont
  {Maslennikov}, \citenamefont {Habl\"utzel}, \citenamefont {Loh},\ and\
  \citenamefont {Matsukevich}}]{Ding2017b}%
  \BibitemOpen
  \bibfield  {author} {\bibinfo {author} {\bibfnamefont {S.}~\bibnamefont
  {Ding}}, \bibinfo {author} {\bibfnamefont {G.}~\bibnamefont {Maslennikov}},
  \bibinfo {author} {\bibfnamefont {R.}~\bibnamefont {Habl\"utzel}}, \bibinfo
  {author} {\bibfnamefont {H.}~\bibnamefont {Loh}}, \ and\ \bibinfo {author}
  {\bibfnamefont {D.}~\bibnamefont {Matsukevich}},\ }\href {\doibase
  10.1103/PhysRevLett.119.150404} {\bibfield  {journal} {\bibinfo  {journal}
  {Phys. Rev. Lett.}\ }\textbf {\bibinfo {volume} {119}},\ \bibinfo {pages}
  {150404} (\bibinfo {year} {2017}{\natexlab{b}})}\BibitemShut {NoStop}%
\bibitem [{\citenamefont {Skelt}\ \emph {et~al.}(2019)\citenamefont {Skelt},
  \citenamefont {Zawadzki},\ and\ \citenamefont {D'Amico}}]{Skelt19}%
  \BibitemOpen
  \bibfield  {author} {\bibinfo {author} {\bibfnamefont {A.~H.}\ \bibnamefont
  {Skelt}}, \bibinfo {author} {\bibfnamefont {K.}~\bibnamefont {Zawadzki}}, \
  and\ \bibinfo {author} {\bibfnamefont {I.}~\bibnamefont {D'Amico}},\ }\href
  {\doibase 10.1088/1751-8121/ab4fb6} {\bibfield  {journal} {\bibinfo
  {journal} {J. Phys. A: Math. Theor.}\ }\textbf {\bibinfo {volume} {52}},\
  \bibinfo {pages} {485304} (\bibinfo {year} {2019})}\BibitemShut {NoStop}%
\bibitem [{\citenamefont {Roberts}\ and\ \citenamefont
  {Clerk}(2020)}]{Roberts20}%
  \BibitemOpen
  \bibfield  {author} {\bibinfo {author} {\bibfnamefont {D.}~\bibnamefont
  {Roberts}}\ and\ \bibinfo {author} {\bibfnamefont {A.~A.}\ \bibnamefont
  {Clerk}},\ }\href {\doibase 10.1103/PhysRevX.10.021022} {\bibfield  {journal}
  {\bibinfo  {journal} {Phys. Rev. X}\ }\textbf {\bibinfo {volume} {10}},\
  \bibinfo {pages} {021022} (\bibinfo {year} {2020})}\BibitemShut {NoStop}%
\bibitem [{\citenamefont {Zhang}\ \emph {et~al.}(2013)\citenamefont {Zhang},
  \citenamefont {Thingna}, \citenamefont {He}, \citenamefont {Wang},\ and\
  \citenamefont {Li}}]{Zhang_2013}%
  \BibitemOpen
  \bibfield  {author} {\bibinfo {author} {\bibfnamefont {L.}~\bibnamefont
  {Zhang}}, \bibinfo {author} {\bibfnamefont {J.}~\bibnamefont {Thingna}},
  \bibinfo {author} {\bibfnamefont {D.}~\bibnamefont {He}}, \bibinfo {author}
  {\bibfnamefont {J.-S.}\ \bibnamefont {Wang}}, \ and\ \bibinfo {author}
  {\bibfnamefont {B.}~\bibnamefont {Li}},\ }\href {\doibase
  10.1209/0295-5075/103/64002} {\bibfield  {journal} {\bibinfo  {journal}
  {Europhys. Lett.}\ }\textbf {\bibinfo {volume} {103}},\ \bibinfo {pages}
  {64002} (\bibinfo {year} {2013})}\BibitemShut {NoStop}%
\bibitem [{\citenamefont {Helt}\ \emph {et~al.}(2010)\citenamefont {Helt},
  \citenamefont {Yang}, \citenamefont {Liscidini},\ and\ \citenamefont
  {Sipe}}]{Helt10}%
  \BibitemOpen
  \bibfield  {author} {\bibinfo {author} {\bibfnamefont {L.~G.}\ \bibnamefont
  {Helt}}, \bibinfo {author} {\bibfnamefont {Z.}~\bibnamefont {Yang}}, \bibinfo
  {author} {\bibfnamefont {M.}~\bibnamefont {Liscidini}}, \ and\ \bibinfo
  {author} {\bibfnamefont {J.~E.}\ \bibnamefont {Sipe}},\ }\href {\doibase
  10.1364/OL.35.003006} {\bibfield  {journal} {\bibinfo  {journal} {Opt.
  Lett.}\ }\textbf {\bibinfo {volume} {35}},\ \bibinfo {pages} {3006} (\bibinfo
  {year} {2010})}\BibitemShut {NoStop}%
\bibitem [{\citenamefont {Sripakdee}\ and\ \citenamefont
  {Yupapin}(2011)}]{Sripakdee11}%
  \BibitemOpen
  \bibfield  {author} {\bibinfo {author} {\bibfnamefont {C.}~\bibnamefont
  {Sripakdee}}\ and\ \bibinfo {author} {\bibfnamefont {P.~P.}\ \bibnamefont
  {Yupapin}},\ }\href {\doibase https://doi.org/10.1016/j.ijleo.2010.03.019}
  {\bibfield  {journal} {\bibinfo  {journal} {Optik}\ }\textbf {\bibinfo
  {volume} {122}},\ \bibinfo {pages} {535} (\bibinfo {year}
  {2011})}\BibitemShut {NoStop}%
\bibitem [{\citenamefont {Vernon}\ and\ \citenamefont {Sipe}(2015)}]{Vernon15}%
  \BibitemOpen
  \bibfield  {author} {\bibinfo {author} {\bibfnamefont {Z.}~\bibnamefont
  {Vernon}}\ and\ \bibinfo {author} {\bibfnamefont {J.~E.}\ \bibnamefont
  {Sipe}},\ }\href {\doibase 10.1103/PhysRevA.91.053802} {\bibfield  {journal}
  {\bibinfo  {journal} {Phys. Rev. A}\ }\textbf {\bibinfo {volume} {91}},\
  \bibinfo {pages} {053802} (\bibinfo {year} {2015})}\BibitemShut {NoStop}%
\bibitem [{\citenamefont {Kowalewska-Kudłaszyk}\ and\ \citenamefont
  {Chimczak}(2019)}]{Kowalewska19}%
  \BibitemOpen
  \bibfield  {author} {\bibinfo {author} {\bibfnamefont {A.}~\bibnamefont
  {Kowalewska-Kudłaszyk}}\ and\ \bibinfo {author} {\bibfnamefont
  {G.}~\bibnamefont {Chimczak}},\ }\href {\doibase 10.3390/sym11081023}
  {\bibfield  {journal} {\bibinfo  {journal} {Symmetry}\ }\textbf {\bibinfo
  {volume} {11}},\ \bibinfo {pages} {1023} (\bibinfo {year}
  {2019})}\BibitemShut {NoStop}%
\bibitem [{\citenamefont {Menotti}\ \emph {et~al.}(2019)\citenamefont
  {Menotti}, \citenamefont {Morrison}, \citenamefont {Tan}, \citenamefont
  {Vernon}, \citenamefont {Sipe},\ and\ \citenamefont {Liscidini}}]{Menotti19}%
  \BibitemOpen
  \bibfield  {author} {\bibinfo {author} {\bibfnamefont {M.}~\bibnamefont
  {Menotti}}, \bibinfo {author} {\bibfnamefont {B.}~\bibnamefont {Morrison}},
  \bibinfo {author} {\bibfnamefont {K.}~\bibnamefont {Tan}}, \bibinfo {author}
  {\bibfnamefont {Z.}~\bibnamefont {Vernon}}, \bibinfo {author} {\bibfnamefont
  {J.~E.}\ \bibnamefont {Sipe}}, \ and\ \bibinfo {author} {\bibfnamefont
  {M.}~\bibnamefont {Liscidini}},\ }\href {\doibase
  10.1103/PhysRevLett.122.013904} {\bibfield  {journal} {\bibinfo  {journal}
  {Phys. Rev. Lett.}\ }\textbf {\bibinfo {volume} {122}},\ \bibinfo {pages}
  {013904} (\bibinfo {year} {2019})}\BibitemShut {NoStop}%
\bibitem [{\citenamefont {Boyd}(2003)}]{boyd03}%
  \BibitemOpen
  \bibfield  {author} {\bibinfo {author} {\bibfnamefont {R.~W.}\ \bibnamefont
  {Boyd}},\ }\href@noop {} {\emph {\bibinfo {title} {Nonlinear optics}}}\
  (\bibinfo  {publisher} {Elsevier},\ \bibinfo {year} {2003})\BibitemShut
  {NoStop}%
\bibitem [{\citenamefont {Gottfried}\ and\ \citenamefont
  {Yan}(2003)}]{Gottfried03}%
  \BibitemOpen
  \bibfield  {author} {\bibinfo {author} {\bibfnamefont {K.}~\bibnamefont
  {Gottfried}}\ and\ \bibinfo {author} {\bibfnamefont {T.-M.}\ \bibnamefont
  {Yan}},\ }\href@noop {} {\emph {\bibinfo {title} {Quantum mechanics:
  fundamentals}}}\ (\bibinfo  {publisher} {Springer Science \& Business
  Media},\ \bibinfo {year} {2003})\BibitemShut {NoStop}%
\bibitem [{\citenamefont {Bosman}\ \emph {et~al.}(2017)\citenamefont {Bosman},
  \citenamefont {Gely}, \citenamefont {Singh}, \citenamefont {Bruno},
  \citenamefont {Bothner},\ and\ \citenamefont {Steele}}]{Bosman17}%
  \BibitemOpen
  \bibfield  {author} {\bibinfo {author} {\bibfnamefont {S.~J.}\ \bibnamefont
  {Bosman}}, \bibinfo {author} {\bibfnamefont {M.~F.}\ \bibnamefont {Gely}},
  \bibinfo {author} {\bibfnamefont {V.}~\bibnamefont {Singh}}, \bibinfo
  {author} {\bibfnamefont {A.}~\bibnamefont {Bruno}}, \bibinfo {author}
  {\bibfnamefont {D.}~\bibnamefont {Bothner}}, \ and\ \bibinfo {author}
  {\bibfnamefont {G.~A.}\ \bibnamefont {Steele}},\ }\href {\doibase
  10.1038/s41534-017-0046-y} {\bibfield  {journal} {\bibinfo  {journal} {npj
  Quantum Inf.}\ }\textbf {\bibinfo {volume} {3}},\ \bibinfo {pages} {46}
  (\bibinfo {year} {2017})}\BibitemShut {NoStop}%
\bibitem [{\citenamefont {Kockum}\ \emph {et~al.}(2019)\citenamefont {Kockum},
  \citenamefont {Miranowicz}, \citenamefont {De~Liberato}, \citenamefont
  {Savasta},\ and\ \citenamefont {Nori}}]{FriskKockum19}%
  \BibitemOpen
  \bibfield  {author} {\bibinfo {author} {\bibfnamefont {A.~F.}\ \bibnamefont
  {Kockum}}, \bibinfo {author} {\bibfnamefont {A.}~\bibnamefont {Miranowicz}},
  \bibinfo {author} {\bibfnamefont {S.}~\bibnamefont {De~Liberato}}, \bibinfo
  {author} {\bibfnamefont {S.}~\bibnamefont {Savasta}}, \ and\ \bibinfo
  {author} {\bibfnamefont {F.}~\bibnamefont {Nori}},\ }\href {\doibase
  10.1038/s42254-018-0006-2} {\bibfield  {journal} {\bibinfo  {journal} {Nat.
  Rev. Phys.}\ }\textbf {\bibinfo {volume} {1}},\ \bibinfo {pages} {19}
  (\bibinfo {year} {2019})}\BibitemShut {NoStop}%
\bibitem [{\citenamefont {Higgs}(1979)}]{Higgs79}%
  \BibitemOpen
  \bibfield  {author} {\bibinfo {author} {\bibfnamefont {P.~W.}\ \bibnamefont
  {Higgs}},\ }\href {\doibase 10.1088/0305-4470/12/3/006} {\bibfield  {journal}
  {\bibinfo  {journal} {J. Phys. A: Math. Gen.}\ }\textbf {\bibinfo {volume}
  {12}},\ \bibinfo {pages} {309} (\bibinfo {year} {1979})}\BibitemShut
  {NoStop}%
\bibitem [{\citenamefont {Sklyanin}(1982)}]{Sklyanin82}%
  \BibitemOpen
  \bibfield  {author} {\bibinfo {author} {\bibfnamefont {E.~K.}\ \bibnamefont
  {Sklyanin}},\ }\href {\doibase 10.1007/BF01077848} {\bibfield  {journal}
  {\bibinfo  {journal} {Funct. Anal. Appl.}\ }\textbf {\bibinfo {volume}
  {16}},\ \bibinfo {pages} {263} (\bibinfo {year} {1982})}\BibitemShut
  {NoStop}%
\bibitem [{\citenamefont {Karassiov}\ and\ \citenamefont
  {Klimov}(1994)}]{Karassiov94}%
  \BibitemOpen
  \bibfield  {author} {\bibinfo {author} {\bibfnamefont {V.~P.}\ \bibnamefont
  {Karassiov}}\ and\ \bibinfo {author} {\bibfnamefont {A.~B.}\ \bibnamefont
  {Klimov}},\ }\href {\doibase https://doi.org/10.1016/0375-9601(94)90816-8}
  {\bibfield  {journal} {\bibinfo  {journal} {Phys. Lett. A}\ }\textbf
  {\bibinfo {volume} {189}},\ \bibinfo {pages} {43} (\bibinfo {year}
  {1994})}\BibitemShut {NoStop}%
\bibitem [{\citenamefont {Klimov}\ \emph {et~al.}(2002)\citenamefont {Klimov},
  \citenamefont {Romero}, \citenamefont {Delgado},\ and\ \citenamefont {nchez
  Soto}}]{Klimov02}%
  \BibitemOpen
  \bibfield  {author} {\bibinfo {author} {\bibfnamefont {A.~B.}\ \bibnamefont
  {Klimov}}, \bibinfo {author} {\bibfnamefont {J.~L.}\ \bibnamefont {Romero}},
  \bibinfo {author} {\bibfnamefont {J.}~\bibnamefont {Delgado}}, \ and\
  \bibinfo {author} {\bibfnamefont {L.~L.~S.}\ \bibnamefont {nchez Soto}},\
  }\href {\doibase 10.1088/1464-4266/5/1/304} {\bibfield  {journal} {\bibinfo
  {journal} {J. Opt. B: Quantum Semiclassical Opt.}\ }\textbf {\bibinfo
  {volume} {5}},\ \bibinfo {pages} {34} (\bibinfo {year} {2002})}\BibitemShut
  {NoStop}%
\bibitem [{\citenamefont {Bonatsos}\ and\ \citenamefont
  {Daskaloyannis}(1999)}]{BONATSOS99}%
  \BibitemOpen
  \bibfield  {author} {\bibinfo {author} {\bibfnamefont {D.}~\bibnamefont
  {Bonatsos}}\ and\ \bibinfo {author} {\bibfnamefont {C.}~\bibnamefont
  {Daskaloyannis}},\ }\href {\doibase 10.1016/S0146-6410(99)00100-3} {\bibfield
   {journal} {\bibinfo  {journal} {Prog. Part. Nucl. Phys.}\ }\textbf {\bibinfo
  {volume} {43}},\ \bibinfo {pages} {537} (\bibinfo {year} {1999})}\BibitemShut
  {NoStop}%
\bibitem [{\citenamefont {Bonatsos}\ \emph {et~al.}(1995)\citenamefont
  {Bonatsos}, \citenamefont {Kolokotronis},\ and\ \citenamefont
  {Daskaloyannis}}]{Bonatsos95}%
  \BibitemOpen
  \bibfield  {author} {\bibinfo {author} {\bibfnamefont {D.}~\bibnamefont
  {Bonatsos}}, \bibinfo {author} {\bibfnamefont {P.}~\bibnamefont
  {Kolokotronis}}, \ and\ \bibinfo {author} {\bibfnamefont {C.}~\bibnamefont
  {Daskaloyannis}},\ }\href {\doibase 10.1142/S0217732395002362} {\bibfield
  {journal} {\bibinfo  {journal} {Mod. Phys. Lett. A}\ }\textbf {\bibinfo
  {volume} {10}},\ \bibinfo {pages} {2197} (\bibinfo {year}
  {1995})}\BibitemShut {NoStop}%
\bibitem [{\citenamefont {Breuer}\ and\ \citenamefont
  {Petruccione}(2002)}]{Breuer02}%
  \BibitemOpen
  \bibfield  {author} {\bibinfo {author} {\bibfnamefont {H.-P.}\ \bibnamefont
  {Breuer}}\ and\ \bibinfo {author} {\bibfnamefont {F.}~\bibnamefont
  {Petruccione}},\ }\href@noop {} {\emph {\bibinfo {title} {The theory of open
  quantum systems}}}\ (\bibinfo  {publisher} {Oxford University Press},\
  \bibinfo {year} {2002})\BibitemShut {NoStop}%
\bibitem [{\citenamefont {Weiss}(2012)}]{Weiss12}%
  \BibitemOpen
  \bibfield  {author} {\bibinfo {author} {\bibfnamefont {U.}~\bibnamefont
  {Weiss}},\ }\href@noop {} {\emph {\bibinfo {title} {Quantum dissipative
  systems}}}\ (\bibinfo  {publisher} {World scientific},\ \bibinfo {year}
  {2012})\BibitemShut {NoStop}%
\bibitem [{\citenamefont {Peskin}(2018)}]{Peskin18}%
  \BibitemOpen
  \bibfield  {author} {\bibinfo {author} {\bibfnamefont {M.}~\bibnamefont
  {Peskin}},\ }\href@noop {} {\emph {\bibinfo {title} {An introduction to
  quantum field theory}}}\ (\bibinfo  {publisher} {CRC press},\ \bibinfo {year}
  {2018})\BibitemShut {NoStop}%
\bibitem [{\citenamefont {Santos}\ and\ \citenamefont
  {Landi}(2016)}]{Santos16}%
  \BibitemOpen
  \bibfield  {author} {\bibinfo {author} {\bibfnamefont {J.~P.}\ \bibnamefont
  {Santos}}\ and\ \bibinfo {author} {\bibfnamefont {G.~T.}\ \bibnamefont
  {Landi}},\ }\href {\doibase 10.1103/PhysRevE.94.062143} {\bibfield  {journal}
  {\bibinfo  {journal} {Phys. Rev. E}\ }\textbf {\bibinfo {volume} {94}},\
  \bibinfo {pages} {062143} (\bibinfo {year} {2016})}\BibitemShut {NoStop}%
\bibitem [{\citenamefont {Alicki}(1979)}]{Alicki79}%
  \BibitemOpen
  \bibfield  {author} {\bibinfo {author} {\bibfnamefont {R.}~\bibnamefont
  {Alicki}},\ }\href {\doibase 10.1088/0305-4470/12/5/007} {\bibfield
  {journal} {\bibinfo  {journal} {J. Phys. A: Math. Gen.}\ }\textbf {\bibinfo
  {volume} {12}},\ \bibinfo {pages} {L103} (\bibinfo {year}
  {1979})}\BibitemShut {NoStop}%
\bibitem [{\citenamefont {Xuereb}\ \emph {et~al.}(2015)\citenamefont {Xuereb},
  \citenamefont {Imparato},\ and\ \citenamefont {Dantan}}]{Xuereb15}%
  \BibitemOpen
  \bibfield  {author} {\bibinfo {author} {\bibfnamefont {A.}~\bibnamefont
  {Xuereb}}, \bibinfo {author} {\bibfnamefont {A.}~\bibnamefont {Imparato}}, \
  and\ \bibinfo {author} {\bibfnamefont {A.}~\bibnamefont {Dantan}},\ }\href
  {\doibase 10.1088/1367-2630/17/5/055013} {\bibfield  {journal} {\bibinfo
  {journal} {New J. Phys.}\ }\textbf {\bibinfo {volume} {17}},\ \bibinfo
  {pages} {055013} (\bibinfo {year} {2015})}\BibitemShut {NoStop}%
\bibitem [{\citenamefont {Cuansing}\ \emph {et~al.}(2012)\citenamefont
  {Cuansing}, \citenamefont {Li},\ and\ \citenamefont {Wang}}]{Cuansing12}%
  \BibitemOpen
  \bibfield  {author} {\bibinfo {author} {\bibfnamefont {E.~C.}\ \bibnamefont
  {Cuansing}}, \bibinfo {author} {\bibfnamefont {H.}~\bibnamefont {Li}}, \ and\
  \bibinfo {author} {\bibfnamefont {J.-S.}\ \bibnamefont {Wang}},\ }\href
  {\doibase 10.1103/PhysRevE.86.031132} {\bibfield  {journal} {\bibinfo
  {journal} {Phys. Rev. E}\ }\textbf {\bibinfo {volume} {86}},\ \bibinfo
  {pages} {031132} (\bibinfo {year} {2012})}\BibitemShut {NoStop}%
\bibitem [{\citenamefont {Wang}\ \emph {et~al.}(2008)\citenamefont {Wang},
  \citenamefont {Wang},\ and\ \citenamefont {L{\"u}}}]{Wang08NEGF}%
  \BibitemOpen
  \bibfield  {author} {\bibinfo {author} {\bibfnamefont {J.-S.}\ \bibnamefont
  {Wang}}, \bibinfo {author} {\bibfnamefont {J.}~\bibnamefont {Wang}}, \ and\
  \bibinfo {author} {\bibfnamefont {J.~T.}\ \bibnamefont {L{\"u}}},\ }\href
  {\doibase 10.1140/epjb/e2008-00195-8} {\bibfield  {journal} {\bibinfo
  {journal} {Eur. Phys. J. B}\ }\textbf {\bibinfo {volume} {62}},\ \bibinfo
  {pages} {381} (\bibinfo {year} {2008})}\BibitemShut {NoStop}%
\bibitem [{\citenamefont {Thingna}\ \emph {et~al.}(2012)\citenamefont
  {Thingna}, \citenamefont {Garc\'{\i}a-Palacios},\ and\ \citenamefont
  {Wang}}]{Thingna12PRB}%
  \BibitemOpen
  \bibfield  {author} {\bibinfo {author} {\bibfnamefont {J.}~\bibnamefont
  {Thingna}}, \bibinfo {author} {\bibfnamefont {J.~L.}\ \bibnamefont
  {Garc\'{\i}a-Palacios}}, \ and\ \bibinfo {author} {\bibfnamefont {J.-S.}\
  \bibnamefont {Wang}},\ }\href {\doibase 10.1103/PhysRevB.85.195452}
  {\bibfield  {journal} {\bibinfo  {journal} {Phys. Rev. B}\ }\textbf {\bibinfo
  {volume} {85}},\ \bibinfo {pages} {195452} (\bibinfo {year}
  {2012})}\BibitemShut {NoStop}%
\bibitem [{\citenamefont {Wang}\ \emph {et~al.}(2014)\citenamefont {Wang},
  \citenamefont {Agarwalla}, \citenamefont {Li},\ and\ \citenamefont
  {Thingna}}]{Wang14}%
  \BibitemOpen
  \bibfield  {author} {\bibinfo {author} {\bibfnamefont {J.-S.}\ \bibnamefont
  {Wang}}, \bibinfo {author} {\bibfnamefont {B.~K.}\ \bibnamefont {Agarwalla}},
  \bibinfo {author} {\bibfnamefont {H.}~\bibnamefont {Li}}, \ and\ \bibinfo
  {author} {\bibfnamefont {J.}~\bibnamefont {Thingna}},\ }\href {\doibase
  10.1007/s11467-013-0340-x} {\bibfield  {journal} {\bibinfo  {journal} {Front.
  Phys.}\ }\textbf {\bibinfo {volume} {9}},\ \bibinfo {pages} {673} (\bibinfo
  {year} {2014})}\BibitemShut {NoStop}%
\bibitem [{\citenamefont {Kane}\ and\ \citenamefont {Fisher}(1997)}]{Kane97}%
  \BibitemOpen
  \bibfield  {author} {\bibinfo {author} {\bibfnamefont {C.~L.}\ \bibnamefont
  {Kane}}\ and\ \bibinfo {author} {\bibfnamefont {M.~P.~A.}\ \bibnamefont
  {Fisher}},\ }\href {\doibase 10.1103/PhysRevB.55.15832} {\bibfield  {journal}
  {\bibinfo  {journal} {Phys. Rev. B}\ }\textbf {\bibinfo {volume} {55}},\
  \bibinfo {pages} {15832} (\bibinfo {year} {1997})}\BibitemShut {NoStop}%
\bibitem [{\citenamefont {Das}\ \emph {et~al.}(2019)\citenamefont {Das},
  \citenamefont {Misra}, \citenamefont {Pal}, \citenamefont {Sen(De)},\ and\
  \citenamefont {Sen}}]{Das19}%
  \BibitemOpen
  \bibfield  {author} {\bibinfo {author} {\bibfnamefont {S.}~\bibnamefont
  {Das}}, \bibinfo {author} {\bibfnamefont {A.}~\bibnamefont {Misra}}, \bibinfo
  {author} {\bibfnamefont {A.~K.}\ \bibnamefont {Pal}}, \bibinfo {author}
  {\bibfnamefont {A.}~\bibnamefont {Sen(De)}}, \ and\ \bibinfo {author}
  {\bibfnamefont {U.}~\bibnamefont {Sen}},\ }\href {\doibase
  10.1209/0295-5075/125/20007} {\bibfield  {journal} {\bibinfo  {journal}
  {Europhys. Lett.}\ }\textbf {\bibinfo {volume} {125}},\ \bibinfo {pages}
  {20007} (\bibinfo {year} {2019})}\BibitemShut {NoStop}%
\bibitem [{\citenamefont {Kosloff}\ and\ \citenamefont
  {Levy}(2014)}]{Levy2014}%
  \BibitemOpen
  \bibfield  {author} {\bibinfo {author} {\bibfnamefont {R.}~\bibnamefont
  {Kosloff}}\ and\ \bibinfo {author} {\bibfnamefont {A.}~\bibnamefont {Levy}},\
  }\href {\doibase 10.1146/annurev-physchem-040513-103724} {\bibfield
  {journal} {\bibinfo  {journal} {Ann. Rev. Phys. Chem.}\ }\textbf {\bibinfo
  {volume} {65}},\ \bibinfo {pages} {365} (\bibinfo {year} {2014})}\BibitemShut
  {NoStop}%
\bibitem [{\citenamefont {Zhou}\ \emph {et~al.}(2015)\citenamefont {Zhou},
  \citenamefont {Thingna}, \citenamefont {Wang},\ and\ \citenamefont
  {Li}}]{Zhou2015}%
  \BibitemOpen
  \bibfield  {author} {\bibinfo {author} {\bibfnamefont {H.}~\bibnamefont
  {Zhou}}, \bibinfo {author} {\bibfnamefont {J.}~\bibnamefont {Thingna}},
  \bibinfo {author} {\bibfnamefont {J.-S.}\ \bibnamefont {Wang}}, \ and\
  \bibinfo {author} {\bibfnamefont {B.}~\bibnamefont {Li}},\ }\href {\doibase
  10.1103/PhysRevB.91.045410} {\bibfield  {journal} {\bibinfo  {journal} {Phys.
  Rev. B}\ }\textbf {\bibinfo {volume} {91}},\ \bibinfo {pages} {045410}
  (\bibinfo {year} {2015})}\BibitemShut {NoStop}%
\bibitem [{\citenamefont {Mitchison}\ \emph {et~al.}(2016)\citenamefont
  {Mitchison}, \citenamefont {Huber}, \citenamefont {Prior}, \citenamefont
  {Woods},\ and\ \citenamefont {Plenio}}]{Mitchison16}%
  \BibitemOpen
  \bibfield  {author} {\bibinfo {author} {\bibfnamefont {M.~T.}\ \bibnamefont
  {Mitchison}}, \bibinfo {author} {\bibfnamefont {M.}~\bibnamefont {Huber}},
  \bibinfo {author} {\bibfnamefont {J.}~\bibnamefont {Prior}}, \bibinfo
  {author} {\bibfnamefont {M.~P.}\ \bibnamefont {Woods}}, \ and\ \bibinfo
  {author} {\bibfnamefont {M.~B.}\ \bibnamefont {Plenio}},\ }\href {\doibase
  10.1088/2058-9565/1/1/015001} {\bibfield  {journal} {\bibinfo  {journal}
  {Quant. Sci. Tech.}\ }\textbf {\bibinfo {volume} {1}},\ \bibinfo {pages}
  {015001} (\bibinfo {year} {2016})}\BibitemShut {NoStop}%
\bibitem [{Note1()}]{Note1}%
  \BibitemOpen
  \bibinfo {note} {Qualitatively, even when we set different initial states, we
  obtain the negative current unless we begin with the global ground
  state.}\BibitemShut {Stop}%
\bibitem [{\citenamefont {Lee}\ and\ \citenamefont
  {Sadeghpour}(2013)}]{Lee2013}%
  \BibitemOpen
  \bibfield  {author} {\bibinfo {author} {\bibfnamefont {T.~E.}\ \bibnamefont
  {Lee}}\ and\ \bibinfo {author} {\bibfnamefont {H.~R.}\ \bibnamefont
  {Sadeghpour}},\ }\href {\doibase 10.1103/PhysRevLett.111.234101} {\bibfield
  {journal} {\bibinfo  {journal} {Phys. Rev. Lett.}\ }\textbf {\bibinfo
  {volume} {111}},\ \bibinfo {pages} {234101} (\bibinfo {year}
  {2013})}\BibitemShut {NoStop}%
\bibitem [{\citenamefont {Herpich}\ \emph {et~al.}(2018)\citenamefont
  {Herpich}, \citenamefont {Thingna},\ and\ \citenamefont
  {Esposito}}]{HerpichPRX18}%
  \BibitemOpen
  \bibfield  {author} {\bibinfo {author} {\bibfnamefont {T.}~\bibnamefont
  {Herpich}}, \bibinfo {author} {\bibfnamefont {J.}~\bibnamefont {Thingna}}, \
  and\ \bibinfo {author} {\bibfnamefont {M.}~\bibnamefont {Esposito}},\ }\href
  {\doibase 10.1103/PhysRevX.8.031056} {\bibfield  {journal} {\bibinfo
  {journal} {Phys. Rev. X}\ }\textbf {\bibinfo {volume} {8}},\ \bibinfo {pages}
  {031056} (\bibinfo {year} {2018})}\BibitemShut {NoStop}%
\bibitem [{\citenamefont {Laskar}\ \emph {et~al.}(2020)\citenamefont {Laskar},
  \citenamefont {Adhikary}, \citenamefont {Mondal}, \citenamefont {Katiyar},
  \citenamefont {Vinjanampathy},\ and\ \citenamefont {Ghosh}}]{Sai2020}%
  \BibitemOpen
  \bibfield  {author} {\bibinfo {author} {\bibfnamefont {A.~W.}\ \bibnamefont
  {Laskar}}, \bibinfo {author} {\bibfnamefont {P.}~\bibnamefont {Adhikary}},
  \bibinfo {author} {\bibfnamefont {S.}~\bibnamefont {Mondal}}, \bibinfo
  {author} {\bibfnamefont {P.}~\bibnamefont {Katiyar}}, \bibinfo {author}
  {\bibfnamefont {S.}~\bibnamefont {Vinjanampathy}}, \ and\ \bibinfo {author}
  {\bibfnamefont {S.}~\bibnamefont {Ghosh}},\ }\href {\doibase
  10.1103/PhysRevLett.125.013601} {\bibfield  {journal} {\bibinfo  {journal}
  {Phys. Rev. Lett.}\ }\textbf {\bibinfo {volume} {125}},\ \bibinfo {pages}
  {013601} (\bibinfo {year} {2020})}\BibitemShut {NoStop}%
\end{thebibliography}%

\end{document}